\documentclass[a4paper, 11pt]{article}
\usepackage{graphicx}
\usepackage[numbers, sort&compress]{natbib}				
\usepackage{amsmath}
\usepackage{amssymb}   
\usepackage{siunitx}
\usepackage[american]{babel}	
\usepackage{caption}
\usepackage{tabularx}
\usepackage{listings} 
\usepackage{hyperref} 
\usepackage[width=13.6cm, height=21cm]{geometry}
\usepackage{lmodern}
\usepackage{microtype}
\usepackage{setspace}
\usepackage{xcolor}
\usepackage{wasysym}  
\usepackage{booktabs}
\usepackage{todonotes}
\usepackage{authblk}
\usepackage{framed}
\usepackage{subfigure} 

\definecolor{halfgray}{gray}{0.55} 
\definecolor{webgreen}{rgb}{0,.5,0}
\definecolor{webbrown}{rgb}{.6,0,0}
\definecolor{RoyalBlue}{cmyk}{1, 0.50, 0, 0}

\DeclareGraphicsExtensions{.pdf,.png,.jpg}

\hypersetup{%
    colorlinks=false, linktocpage=false, pdfstartpage=1, 
    breaklinks=true, pdfpagemode=UseNone, pageanchor=true, pdfpagemode=UseOutlines,%
    plainpages=false, bookmarksnumbered, bookmarksopen=true, bookmarksopenlevel=1,%
    hypertexnames=true, pdfhighlight=/O,
    urlcolor=webbrown, linkcolor=RoyalBlue, citecolor=webgreen, 
    pdfsubject={},%
    pdfkeywords={},%
    pdfcreator={pdfLaTeX},%
    pdfproducer={LaTeX with hyperref}%
}   

\makeatletter
\@ifpackageloaded{babel}%
    {%
       \addto\extrasamerican{%
				}%
    }{\relax}
\makeatother

\newpage
\title{\Huge{Quantifying the Evolutionary \\  Self Structuring of \\
Embodied Cognitive Networks
}}

\author[1,2]{Fabio Bonsignorio $^*$}

\affil[1]{RoboticsLab
Department of System Engineering and Automation\\
University Carlos III of Madrid\\ Av. Universidad 30, 28911 Leganes (Madrid), Spain\\ Email:fabio.bonsignorio@uc3m.es $|$ +34 91 624 6014
}
\affil[2]{HeronRobots srl\\
V.R.Ceccardi 1/18, 
16121 Genova (Ge), Italy \\ Email: fabio.bonsignorio@heronrobots.com
}

\date{}
\begin{document}
\maketitle $^*$ Corresponding author.
\begin{abstract}
We outline a possible theoretical framework for the quantitative modeling of networked embodied cognitive systems. We notice that: 1) information self structuring through sensory-motor coordination does not deterministically occur in $ \mathbb{R}^n $ vector space, a generic multivariable space, but in SE(3), the 'group' structure of the possible motions of a body in space; 2) it happens in a stochastic open ended environment. These observations may simplify, at the price of a certain abstraction, the modeling and the design of self organization processes based on the maximization of some informational measures, such as mutual information. Furthermore, by providing closed form or computationally lighter algorithms, it may significantly reduce the computational burden of their implementation. 
We propose a modeling framework which aims to give new tools for the design of networks of new artificial self organizing, embodied and intelligent agents and the 'reverse engineering' of natural ones.
At this point, it represents much a theoretical conjecture and it has still to be experimentally verified  whether this model will be useful in practice.
\vspace{5mm}

\noindent
Keywords\\
Self-organization, embodiment, stochasticity, information, Lie groups, theoretical models
\end{abstract}

\newpage

\section{Introduction}\label{sec:intro}

In nature cognitive adaptation is part of the overall adaptation process. On the one hand, a living being consuming less energy, all the other conditions staying the same, is more likely to survive.  On the other hand, in many ecological niches, and for the large majority of animals, the ones which move, the capability of processing more information in less time gives them a definite evolutionary advantage as it allows them to move faster. Although there are other fitness criteria, depending on the species and the environments where they live, there is always a pressure to improve energy efficiency and maximize information processing capability. Moreover 'short' control programs have a higher probability to emerge from a random process in comparison to longer ones. As a consequence, we observe a natural pressure to offload part of the information processing to the system dynamics. The exploitation of 'morphological computation' schemes is a natural way to cope with these evolutionary pressures. These efficiency and effectiveness features, low energy consumption, high 'intelligence', are also needed by artificial intelligent systems: many robots need autonomy also in the energetic sense, and 'simple' controllers are usually more reliable. Natural cognitive processes emerge from loosely coupled networks of embodied biological neurons, see for a detailed discussion of this point ~\cite{Lungarellaetal:2007:50Years,Bonsignorio:2007:Preliminary,Bonsignorio:2009:Steps}.  It has, also, been shown that the successful adaptation of the 
sensory-motor coordination, in a wide set of physical robotics settings, is usually characterized by a peak of the mutual (or multi) information between the sensor and the actuators,  ~\cite{LungarellaSporns:2006:MappingInformationFlow}. In ~\cite{Chirikjian:2010:Information,Bonsignorio:2010:OntheStochastic} it is argued that the recognition of the Lie group structure of the mobility space may help planning methods based on searching in the configurations space. It has recently been shown that this allows us, given the statistical distribution of the joint variables, in certain cases, to analyze the controllability, observability and stability of (some) kinematic chains from a Shannon information standpoint with compact closed-form relations, in ~\cite{Bonsignorio:2010:OntheStochastic}.  Several researchers have shown the importance of Information Driven Self Organization (IDSO), in particular Prokopenko, Der and others,~\cite{Prokopenkoetal:2006:EvolvingSpatiotemporal, Deretal:2006:Letitroll,Ayetal:2008:Predictive},  who used simulations of snake-bots, humanoids and grasping systems. These approaches seem very promising.


The combination of self organization processes based on the maximization of suitable information metrics and the exploitation of the inherent structure of the motion of a macroscopic physical body might enable the design of effective and robust self organized controllers and behaviors for sensory-motor coordination. It remains an open question whether this possibility is exploited or not in nature.


The main contribution of this paper, which is of a rather hypothetical, conceptual and theoretical nature, is to show that it is possible to design self organizing emergent controllers of reduced computational cost by exploiting the robot morphology. This can be done by merging ideas coming from Prokopenko, Der et alias, with differential geometry and stochastic kinematics models developed, among others, by Chirijkian. The resulting method, proposed in this paper, is suitable also for deformable, 'soft', robot systems and might be exploited in nature by animals.

Moreover we review the basic ideas from IDSO, Information Theory, differential geometry and stochastic kinematics, which might play an important role in the development of a quantitative theoretical framework for morphological computation.

The development of significant field experiments will be part of future work.

Despite the intuition and the valid arguments, shared by many researchers and scholars, that in nature some kind of embodied emergent self organization process might be the general organizational principle of cognitive processes and of self and consciousness, the development of a quantitative framework for the modeling (and synthesis) of those processes has not been achieved, so far. As an example we may observe that while the Cornell Ranger,~\cite{Cornellranger:2011:Website}, can walk for tens of Kilometers on a single battery charge, there is no way to change the speed, which depends on the morphology of the system, 'to change speed, you must change the robot'. A soft (variable impedance) legged system may in principle change speed, but how  the controller should be shaped ?How should you ground emergent embodied cognitive processes? How should you quantify the 'performance', or better the 'fitness', of the different design options?


The modeling approach proposed here aims to provide an initial instantiation of a design methodology for the orchestration of emerging self organizing controllers in soft robots and a possible model of similar processes happening in nature.

In what follows, we first review, in section 2, some important results about the informational metrics of sensory-motor coordination, then in section 3 we illustrate the concept of information driven self organization (IDSO) and we discuss a number of examples from the literature. In section 4, we show how the morphology of a physical body affects self organization of information structures in physically embedded agents and how modeling these processes allows the design of more robust self organizing controllers.

In section 5 we analyze the conceptual and theoretical implications of this model and we outline the open challenges and the future work.

We summarize in a number of framed boxes concepts and definitions, that,  while possibly obvious to readers with a background in information theory, differential geometry or stochastic kinematics, might constitute a serious obstacle for the understanding of others.

The mathematical details are described in the appendices.

\begin{framed}
\begin{center}Basic Information Metrics\end{center}

 The (Shannon) entropy offers a way to quantify the 'information' carried by a stochastic variable with an associate probability distribution, p(x):
\begin{equation}
H\left( x \right) = \sum\limits_{x \in X} {{p_x}\left( x \right)} \ln {p_x}\left( x \right)\label {eq:xdef}\end{equation}
It can be proven that any metrics with the reasonable properties we want for an information metric must have a similar form.
In particular let us consider some reasonable properties like:
\begin{enumerate}
\item{\bf Continuity}
The metric should be  such that if you change  the values of the probabilities by a very small amount , the metric should only change by only a small amount.
\item{\bf Symmetry}
The metric should not change if the results $ {x_i} $ are re-ordered, i.e.:
\begin{equation}
{H_n}\left( {{p_x\left( x_1 \right)},{p_x\left( x_2 \right)},...,{p_x\left( x_n \right)}} \right) = {H_n}\left( {{p_x\left( x_2 \right)},{p_x\left( x_1 \right)},...,{p_x\left( x_n \right)}} \right)
\label{eq:xdef1}\end{equation}

\item{\bf  Maximum}
The metric should be maximal if all the outcomes are equally likely (uncertainty is highest when all possible events are equiprobable) 
\begin{equation}
{H_n}\left( {{p_x\left( x_1 \right)},...,{p_x\left( x_n \right)}} \right) \le {H_n}\left( {\frac{1}{n},...,\frac{1}{n}} \right)
\label
{eq:xdef2}\end{equation}

For equiprobable events the entropy should increase with the number of outcomes:
\begin{equation}
\underbrace{{H_n}\left( {\frac{1}{n},...,\frac{1}{n}} \right)}_{\mbox{ $n$}} <\underbrace{ {H_{n + 1}}\left( {\frac{1}{{n + 1}},...,\frac{1}{{n + 1}}} \right)}_{\mbox{ $n+1$}}\label
{eq:xdef3}\end{equation}
\item{\bf Additivity}
The amount of entropy should be independent of how the process is regarded as being divided into parts

If a functional 'H' has these properties, then it must have a form similar to that in equation \eqref{eq:xdef}, i.e.: 
\begin{equation}
 - K \sum\limits_{x \in X} {{p_x}\left( x \right)} \ln {p_x}\left( x \right)
\label
{eq:xdef4}\end{equation}
\end{enumerate}

In a sense Shannon entropy is the simpler way (K=1) to associate a value to the information content of a stochastic variable. This makes it an interesting tool for the study of behaviors in an uncertain world.
Another useful metric, as it can be used to evaluate control systems in terms of how much they contribute to reduce the uncertainty on the state of the controlled variable,~\cite{TouchetteLloyd:2003:Informationtheoreticapproach,Bonsignorio:2007:Preliminary,Bonsignorio:2009:Steps}, is given by 'Mutual Information'.
The mutual information between two given variables is given by equation \eqref{eq:xdef5}, where X and Y are two random variables:

\begin{equation}
I\left( {X,Y} \right) =  - \sum\limits_{x \in X} {\sum\limits_{y \in Y} {{p_{xy}}} } \left( {x,y} \right)\ln \frac{{{p_x}\left( x \right){p_y}\left( y \right)}}{{{p_{xy}}\left( {x,y} \right)}}
\label
{eq:xdef5}\end{equation}

If X and Y are statistically independent the equation above  gives  I(X,Y)=0 ('X', capital letter, represents the set of all the 'x' values and 'Y', capital letter, represents the set of all the 'y' values).
A reference text on these and other topics related to Information Theory is \emph{Elements of Information Theory}, by Cover and Thomas, ~\cite{CoverThomas:2006:Elements}.
\end{framed}

\section{Informational Measures of Sensory-motor Coordination}
\begin{figure}[!ht]
\centering
\includegraphics[width=\textwidth]{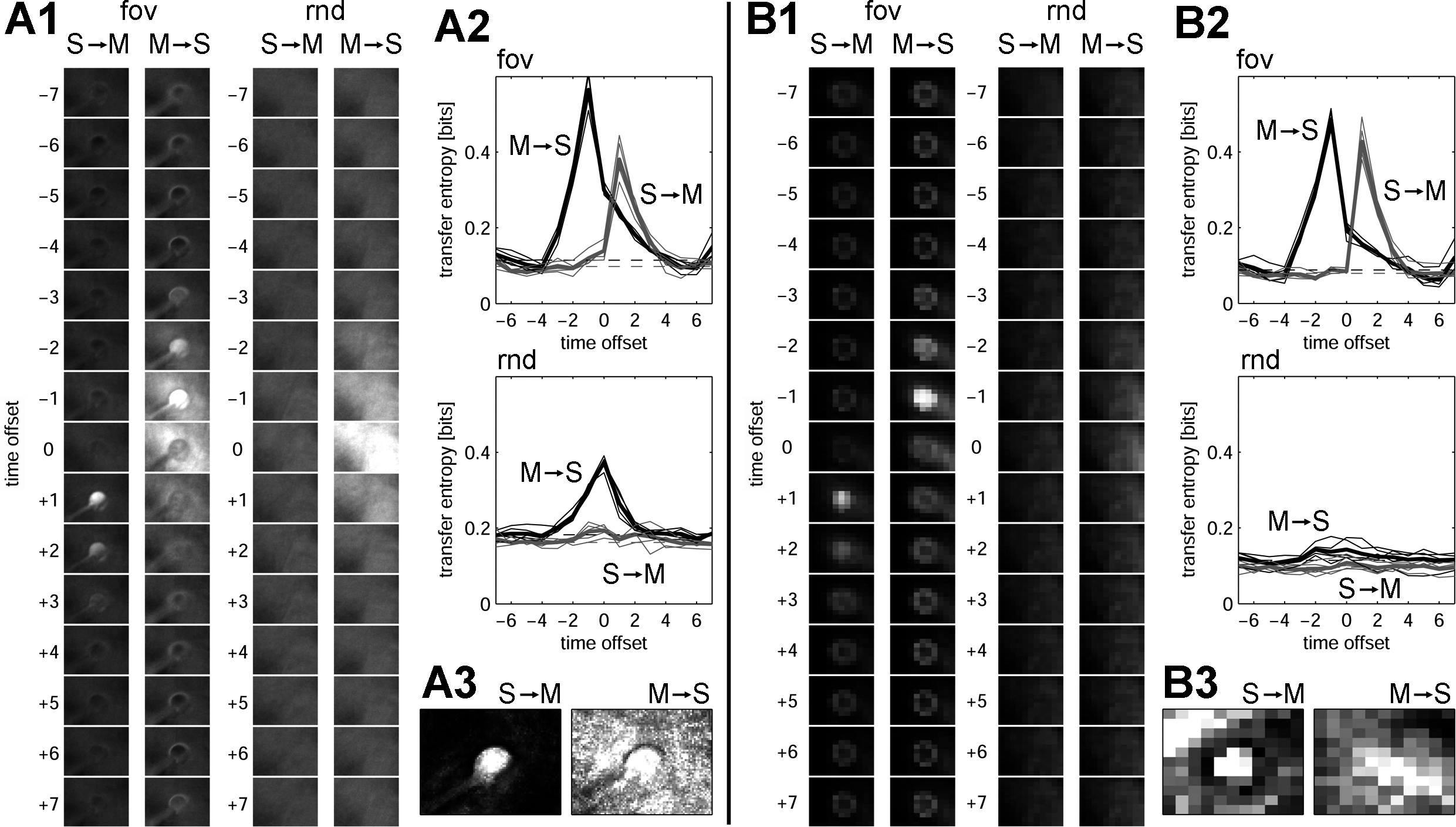}      
   \caption{Evolution of some sensory-motor metrics for a number of physical settings, from ~\cite{LungarellaSporns:2006:MappingInformationFlow}. For all the three different experimental settings considered there is a peak in transfer entropy when a proper sensory-motor coordination is achieved.(Courtesy of the authors)}
  \label{Fig:Evosensorymotor}
\end{figure}

\begin{figure}[!ht]
   \centering
      \includegraphics[width=\linewidth]{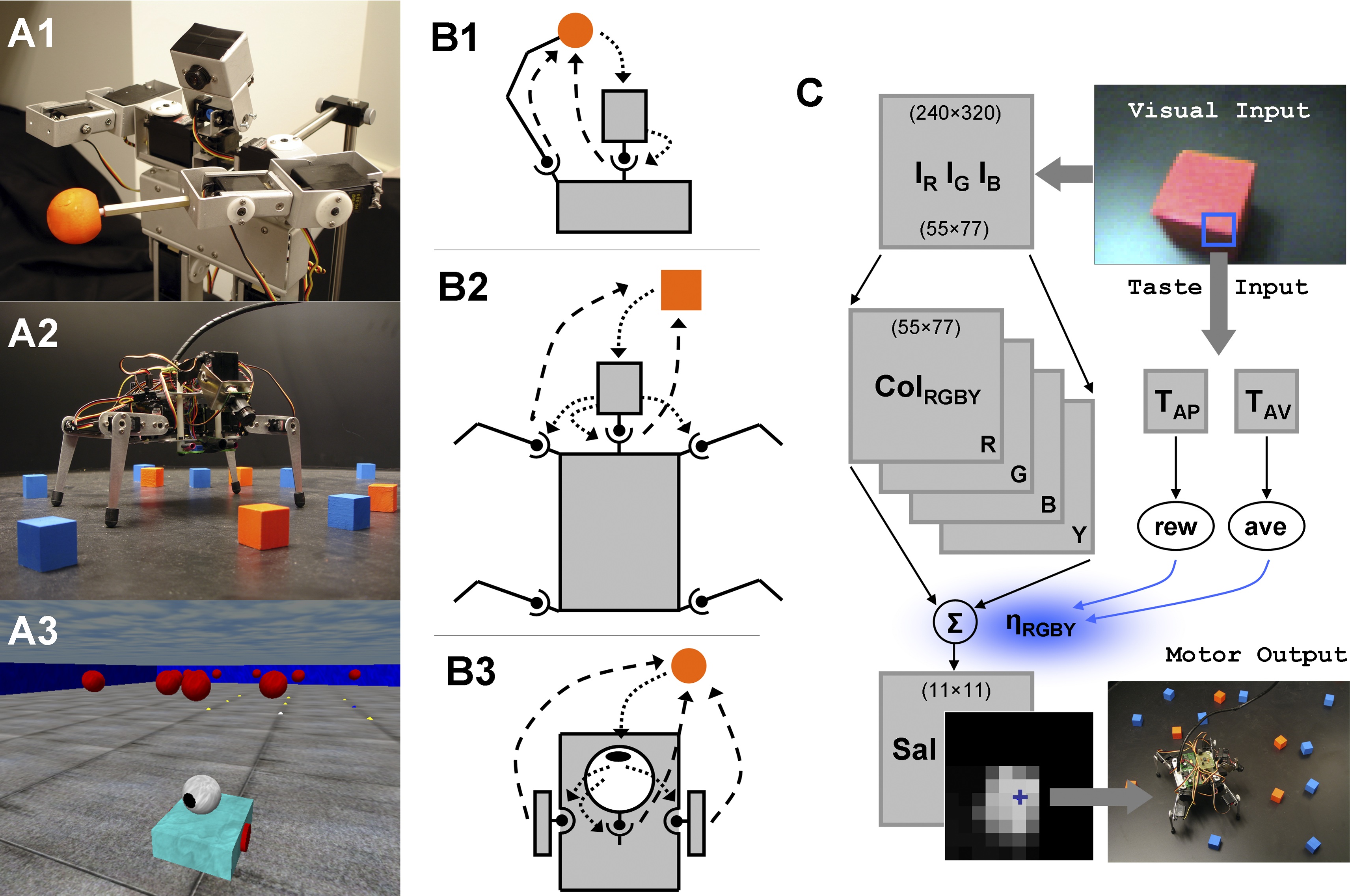}
   \caption{Experimental settings considered in ~\cite{LungarellaSporns:2006:MappingInformationFlow}. A1: a simple humanoid with an eye and an arm manipulating a ball. A2: a spider robot manipulating colored cubes, and A3: a differential wheeled robot with a pan and tilt camera.(Courtesy of the authors)}
  \label{Fig:Expsettings}
\end{figure}

If we consider the analysis in  ~\cite{TouchetteLloyd:2003:Informationtheoreticapproach}, which shows how greater values of mutual information between the sensors and the actuators characterize 'good' controllers, the results in ~\cite{LungarellaSporns:2006:MappingInformationFlow}, summarized in Figures ~\ref{Fig:Evosensorymotor} and ~\ref{Fig:Expsettings},  are well understood.  Lungarella and Sporns have applied some information metrics related to Shannon entropy and mutual information to qunatitatively characterize the sensory-motor coordination of a set of physically different physical agents. 
\vspace{0.5cm}
\begin{center}
[Insert Figure~\ref{Fig:Evosensorymotor}]
\end{center}
\vspace{0.5cm}
\vspace{0.5cm}
\begin{center}
[Insert Figure~\ref{Fig:Expsettings}]
\end{center}
\vspace{0.5cm}

\begin{figure}[!ht]
   \centering
      \includegraphics[width=0.75\linewidth]{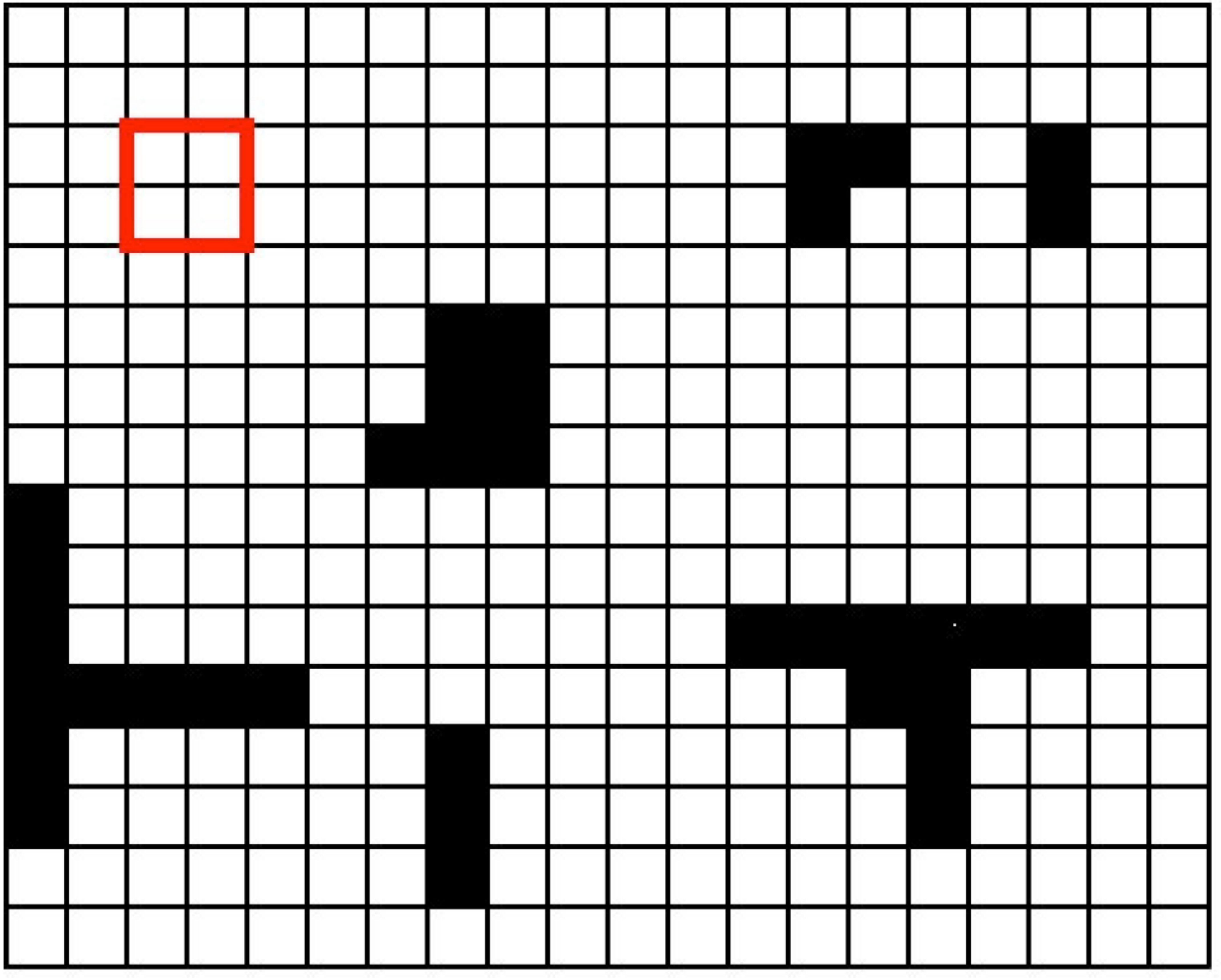}
   \caption{How to calculate the entropy of a cluttered environment, from  ~\cite{LampeChatila:2006:Performance}. The red square defines the 'atomic' cell on which the density of obstacles is calculated. The density of obstacles is interpreted as a probability of finding an obstacle and integrated to give the Shannon entropy of the obstacle distribution given by equation~\eqref{eq:xdef6}. }
  \label{Fig:Clutteredenvs}
\end{figure}

In Figure~\ref{Fig:Expsettings} the 'information flow' (transfer entropy) between sensory input, neural representation of saliency, and the actuator variables in the various physical settings is represented.
Transfer entropy, like Granger's causality, is a  way to measure by which amount the future value of a time series is related to the current value of another, in some sense it quantifies how strongly a dynamic variable depends on another one.
As mentiones earlier, transfer entropy peaks when the system has learned an effective sensorimotor coordination schema (materialized by the weights in the neural networks).
The 'environments' used in ~\cite{LungarellaSporns:2006:MappingInformationFlow} are intuitively very simple and we would like to run experiments in more complex ones. What we need is a way to characterize the complexity of environments.
But how to quantify how much 'cumbersome' an environment is?
Lampe and Chatila (2006),~\cite{LampeChatila:2006:Performance}, have proposed to measure the complexity of a simple environment by defining a metric based on Shannon Entropy.
With reference to Figure~\ref{Fig:Clutteredenvs}, $ H $ is defined as the entropy related to the density of obstacles:
 \begin{equation}
H = \sum\limits_i {p\left( {{d_i}} \right)} \log p\left( {{d_i}} \right)
\label
{eq:xdef6}\end{equation}

\vspace{0.5cm}
\begin{center}
[Insert Figure~\ref{Fig:Clutteredenvs}]
\end{center}
\vspace{0.5cm}



Where $ p\left( {{d_i}} \right) $ represents the $ {i^{th}} $ density level in the occupancy grid, represented by the red square, with:
\begin{equation}
\sum\limits_i {p\left( {{d_i}} \right)}  = 1
\label
{eq:xdef7}\end{equation}

This metric seems to capture what we intuitively mean when we say that an environment is cluttered, as it gives higher entropy when the grid  cells are occupied in a more random way.

We would, ideally, like to be able to understand a priori for a given environment the minimum complexity of an agent embodied 'brain' and system morphology dynamics to be able to perform a given set of tasks in that environment, relating this to its 'complexity'. This might be  connected to what Ashby called the 'principle of requisite variety',~\cite{Ashby:1960:Design}. 
The aim of the principles and the mathematical tools provide in this paper is to go beyond a purely 'informational' view and devise ways to include the morphology, the dynamics and implicitly the materials into the model of the processes involved by the interaction of an embodied intelligent agent in its environment.
\newpage

\section{Information Driven Self Organization}

The observations made above and the experiments such as those summarized in the previous section suggest that the maximization of some metrics built on mutual information might be an important mechanism at work also in natural intelligent systems. The optimization of suitable informational measures might guide the emergence of cognitive processes and intelligent behavior in animals: this is what we refer to when we talk of Information Driven Self Organization.    Information metrics, as we have seen in previous section, are useful to study complex, in particular multiagent, information processing systems. In what Prokopenko calls a 'strong' form: IDSO could be regarded as one of the main drivers of natural evolution. Snakebot by Tanev,~\cite{Prokopenkoetal:2006:EvolvingSpatiotemporal,Tanevetal:2005:Automated}, is an example of a system designed according to IDSO principles. It is interesting as it shows that 'lifelike' movements emerge by maximizing suitable information metrics.
Another example is given by the hexapod walking model proposed by Cruse,~\cite{Cruse:1996:Neural}, where walking behavior emerges, without any central controller, through the interaction of the embodied system with the environment.
The basic idea is that the parameters of a quite generic control system are tuned to maximize the meaningful interaction of the physical agent with then environment, and that the measure of this coupling is given by, for example, predictive information.

\begin{figure}[!ht]
\centering
\mbox{\subfigure{(a)}{\includegraphics[width=0.48\linewidth]{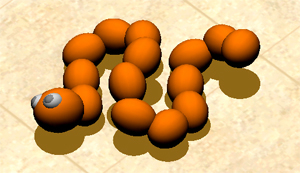}}\quad
\subfigure{(b)}{\includegraphics[width=0.32\linewidth] {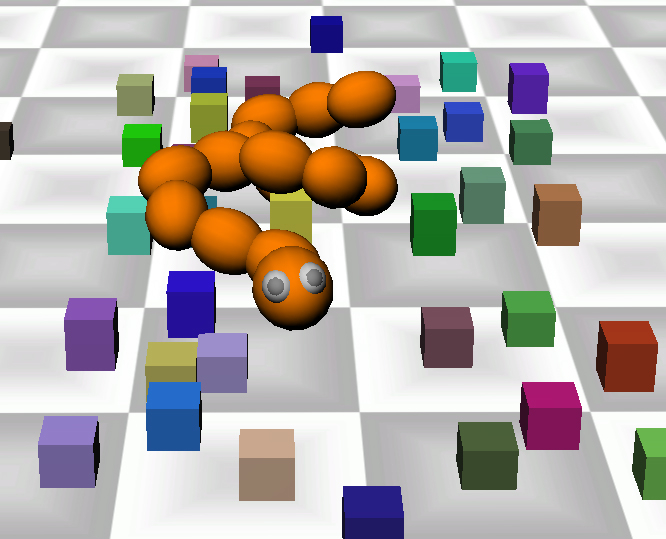}}}
\caption{Snakebot by Tanev.(a) Shows the snakebot, depicted as a series of loosely coupled balls, moving on a plane (b) Shows it while moving in between a number of obstacles, from ~\cite{Tanevetal:2005:Automated}.(Courtesy of the authors)} \label{Fig:Snakebot}
\end{figure}

\vspace{0.5cm}
\begin{center}
[Insert Figure~\ref{Fig:Snakebot}]
\end{center}
\vspace{0.5cm}
  
The 'snakebot' is a simplified model of a rattlesnake, consisting of a series of loosely connected balls.
It can be shown that the amount of predictive information between groups of actuators (measured via generalized excess entropy) grows as the modular robot starts to move across the terrain. The distributed actuators become more coupled when a coordinated side-winding locomotion becomes dominant. Note that if the states of the remote segments are synchronized then some information has been indirectly transferred via stigmergy (due to the physical interactions among the segments, and with the terrain). 
As observed above, on the one hand the application of the predictive information maximization as a self organizing method to generate behavior leads to outcomes which look 'realistic'. On the other hand the calculations involved are heavy, raising doubts on how it could happen in nature (and poses challenges to the utilization of these principles in the synthesis of artifacts). 
In ~\cite{Chirikjian:2010:Information,Bonsignorio:2010:OntheStochastic} it is argued that the incorporation of the Lie group structure which characterizes the mobility space of an embodied agent may be helpful for planning methods based on searching in the configurations space. Stated differently: including the body representation in the orchestrating 'controller' simplifies the control.
In this paper a method that, on the basis of the previously quoted results, may help the quantitative modeling of, natural and artificial, networked embodied systems is described. In the next section 4 we show how this method can be applied to the evolution of sensory networks with emerging controllers  and to the examples in section 2, while in section 5 we discuss how this approach might be generalized to more challenging modeling contexts.
In the paragraph 3.1 we review as an example of IDSO, a model of the evolution of sensory layouts in embodied agents, based on purely informational methods. We will reconsider this example when showing how to incorporate the 'morphology', through Lie groups, in the scheme.

\subsection{Evolution of Sensory Layouts: Ashby's proposal}
Let us consider an example where an IDSO approach is used to evolve a morphology, in particular the morphological distribution of sensors in an embodied agent. It is labeled as Ashby's proposal because the IDSO algorithms perform a pure optimization of informational metrics with no explicit condideration of the embodiment of the systems.The results are interesting as they show that different morphologies of the sensors have different information processing efficiencies, i.e. morphology matters.
In ~\cite{Olssonetal:2004:InformationTrade-Offs}  a microbial genetic algorithm (GA) (taken from Harvey, ~\cite{Harvey:2001:Artificialevolution, McGregorHarvey:2005:Embracing Plagiarism}) was used to evolve a sensory layout. 
The Microbial Genetic Algorithm is a bioinspired evolutionary algorithm that mimics the way microbes exchange DNA between different members of the population, 'horizontally' as opposed to 'vertically' from generation to generation. It uses a 'steady state' method rather than a 'generational' method, meaning that instead of accumulating a complete new generation of offspring, and then eliminating all the members of the older generation and replacing it completely by the new, a single new offspring is generated at a time; then (in order to keep the population size constant) one member of the population dies and it is replaced by the new one.  The selection criterion can be implemented, either by applying a fitness criterion to choose which parents will have an offspring, or by choosing which individual will die. The main benefit of the 'steady state' over the 'generational' criteria are that they are usually easier to implement and that they can be implemented in parallel. In summary the Microbial Genetic Algorithm works as follow: 
\begin{enumerate}
\item pick  two members of the population at random to be parents of the new offspring
\item the least fit of the two parents is chosen as the one to die and be replaced. 
\item information can be transmitted 'horizontally' within a generation
\end{enumerate}

\begin{figure}
\centering
\begin{tabular}{cc}
\subfigure{\includegraphics[width=0.4\linewidth]{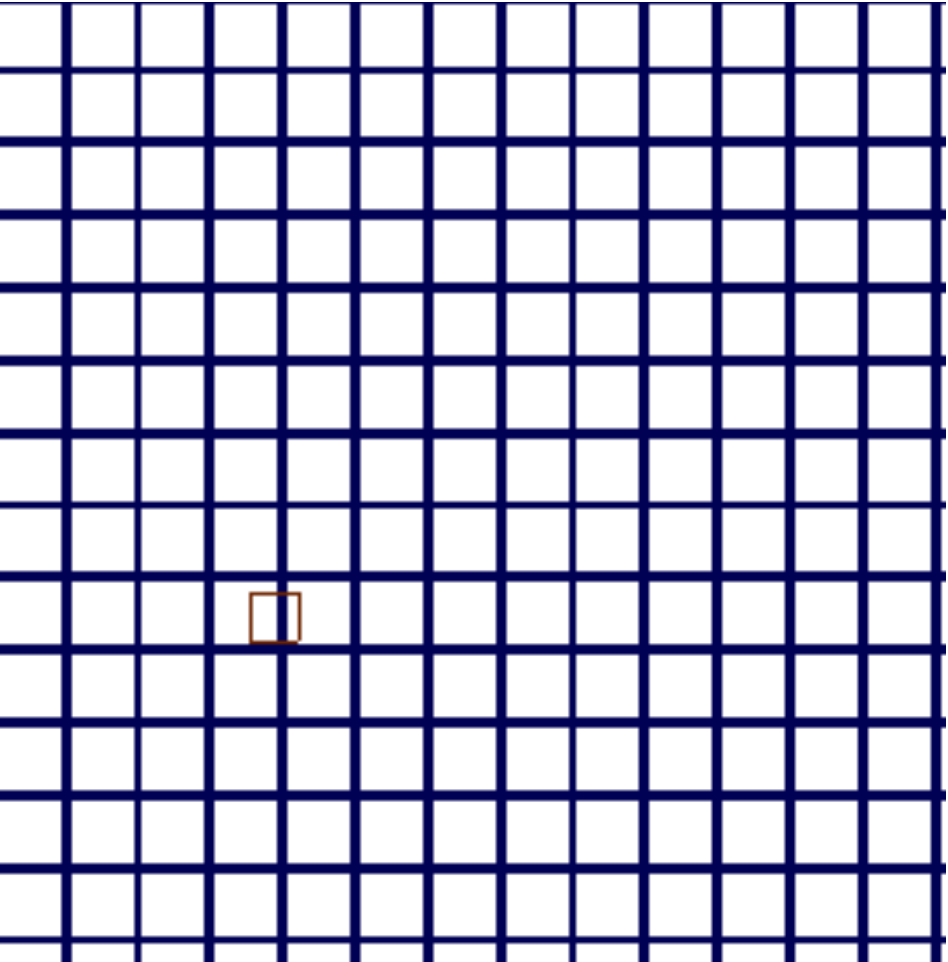} } & 
\subfigure{\includegraphics[width=0.4\linewidth]{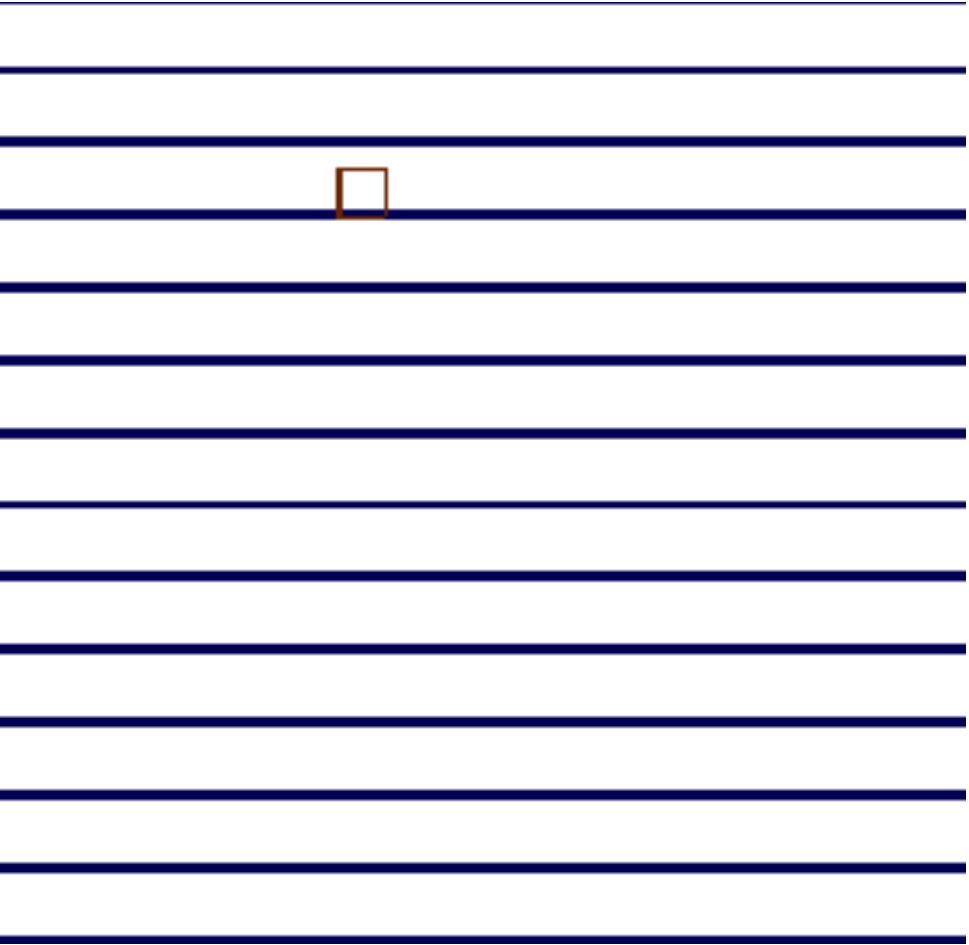} } \\
\subfigure{\includegraphics[width=0.4\linewidth]{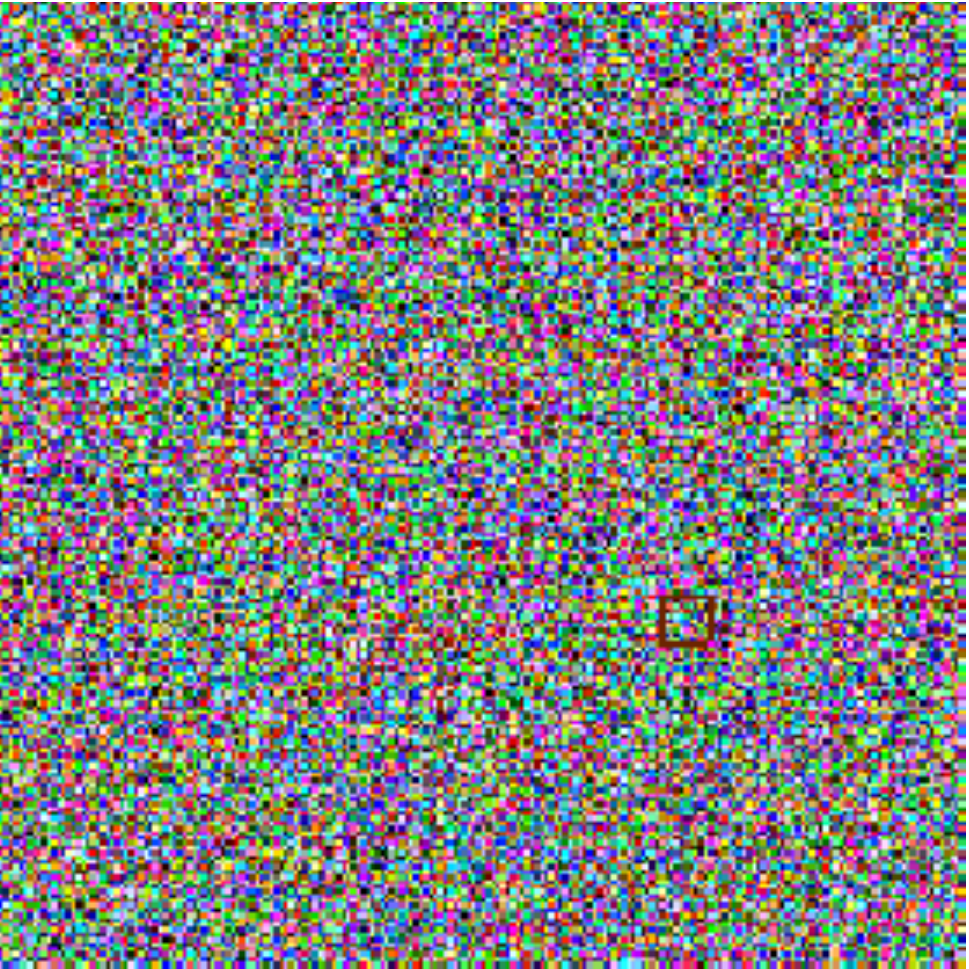} } &
\subfigure{\includegraphics[width=0.4\linewidth]{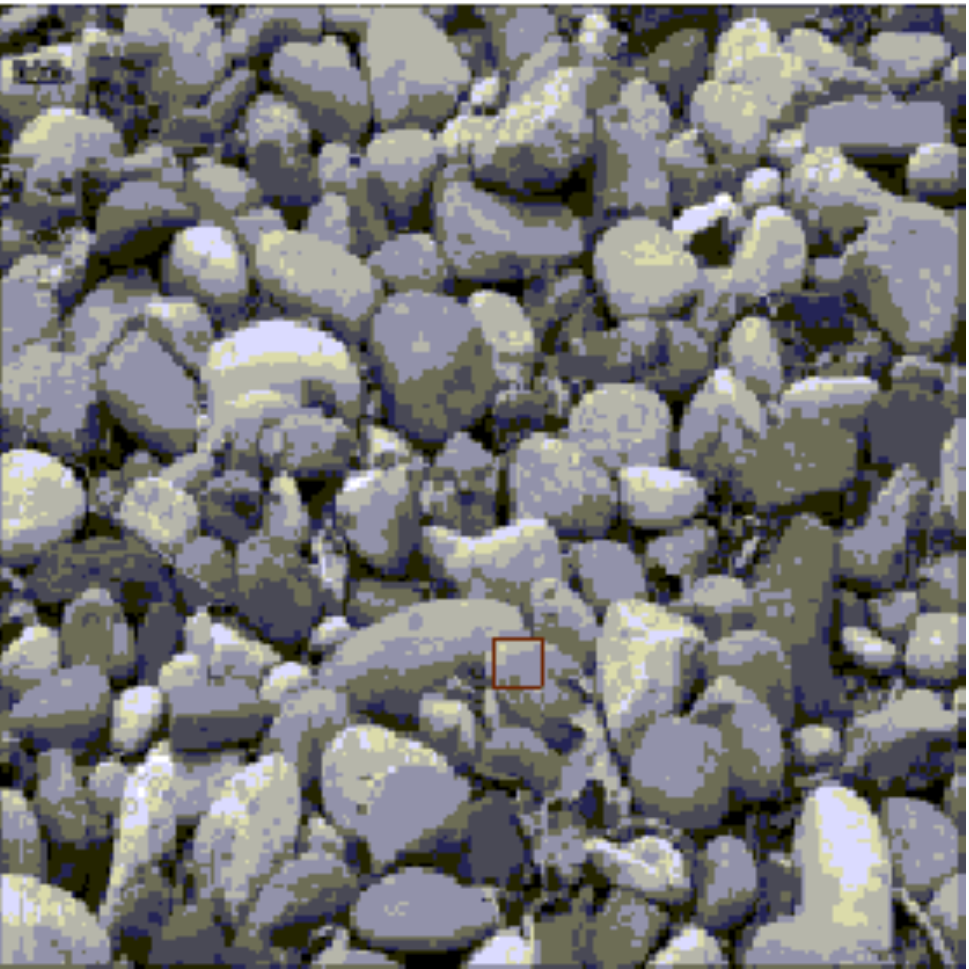} }
\end{tabular}\caption{The different environments used in ~\cite{Olssonetal:2004:InformationTrade-Offs} to evolve the sensor layout.The environments must be sensed by an array of visual sensors, Theit geometric distribution on a flat 10 by 10 square is optimized in order to maximixe their efficiency. The environment is modeled by a grid, a series of horinzontal lines, a noisy colored backgrouns and a picture representing real stones. The red square is used as the 'atomic' cell where to calculate the metrics like in Figure~\ref{Fig:Clutteredenvs}.(Courtesy of the authors)}
  \label{Fig:Environmentsevo}
\end{figure}

Of course any kind of optimizing strategy would fit here, an evolutionary programming kind of strategy might be closer to what happens in nature, but, it is possible that, for example, a simulated annealing or any other kind of optimizing method would work as well.
The task for the evolved sensor layouts is to efficiently and effectively sense the environment which is captured by the fitness function $ ic $, whose expression is given in \eqref{eq:xdef8}.
Each individual's body is modeled by a 10 x 10 square with 10 sensors placed somewhere on it. The genome encodes a list of 10 positions within that square. At any generation the individuals with the sensory layout maximizing a performance index based on informational measures are selected. The evolutionary algorithm has been tested on a number of different simplified environments, see Figure~\ref{Fig:Environmentsevo}.  
The 'fitness' function is given by a weighted performance index:
\begin{equation}
ic(S) = \sum\limits_{X \in S} {\sum\limits_{Y \in S} {\left( {{w_{mi}}I\left( {X;Y} \right) + {w_{cim}}\left( {H\left( {X\left| Y \right.} \right) + H\left( {Y\left| X \right.} \right)} \right)} \right)} } 
\label
{eq:xdef8}\end{equation}
This performance index balances 'redundancy'  through the mutual information between the sensors, 
\begin{equation}
I\left( {X;Y} \right) = H\left( X \right) - H\left( {X\left| Y \right.} \right) = H\left( Y \right) - H\left( {Y\left| X \right.} \right)
\label
{eq:xdef9}\end{equation}
and novelty through the Crutchfield's information metric
\begin{equation}
d\left( {X,Y} \right) = H\left( {X\left| Y \right.} \right) + H\left( {Y\left| X \right.} \right)
\label
{eq:xdef10}\end{equation}
where the conditional entropy $ \ H(Y\left| X \right.)\ $ is expressed as:
\begin{equation}
H\left( {Y\left| X \right.} \right) =  - \sum\limits_{x \in X} {\sum\limits_{y \in Y} {p\left( {x,y} \right)} } \log p\left( {y\left| x \right.} \right)
\label
{eq:xdef11}\end{equation}

\vspace{0.5cm}
\begin{center}
[Insert Figure~\ref{Fig:Environmentsevo}]
\end{center}
\vspace{0.5cm}



\begin{figure}
\centering
\begin{tabular}{cc}
\subfigure{\includegraphics[width=0.4\linewidth]{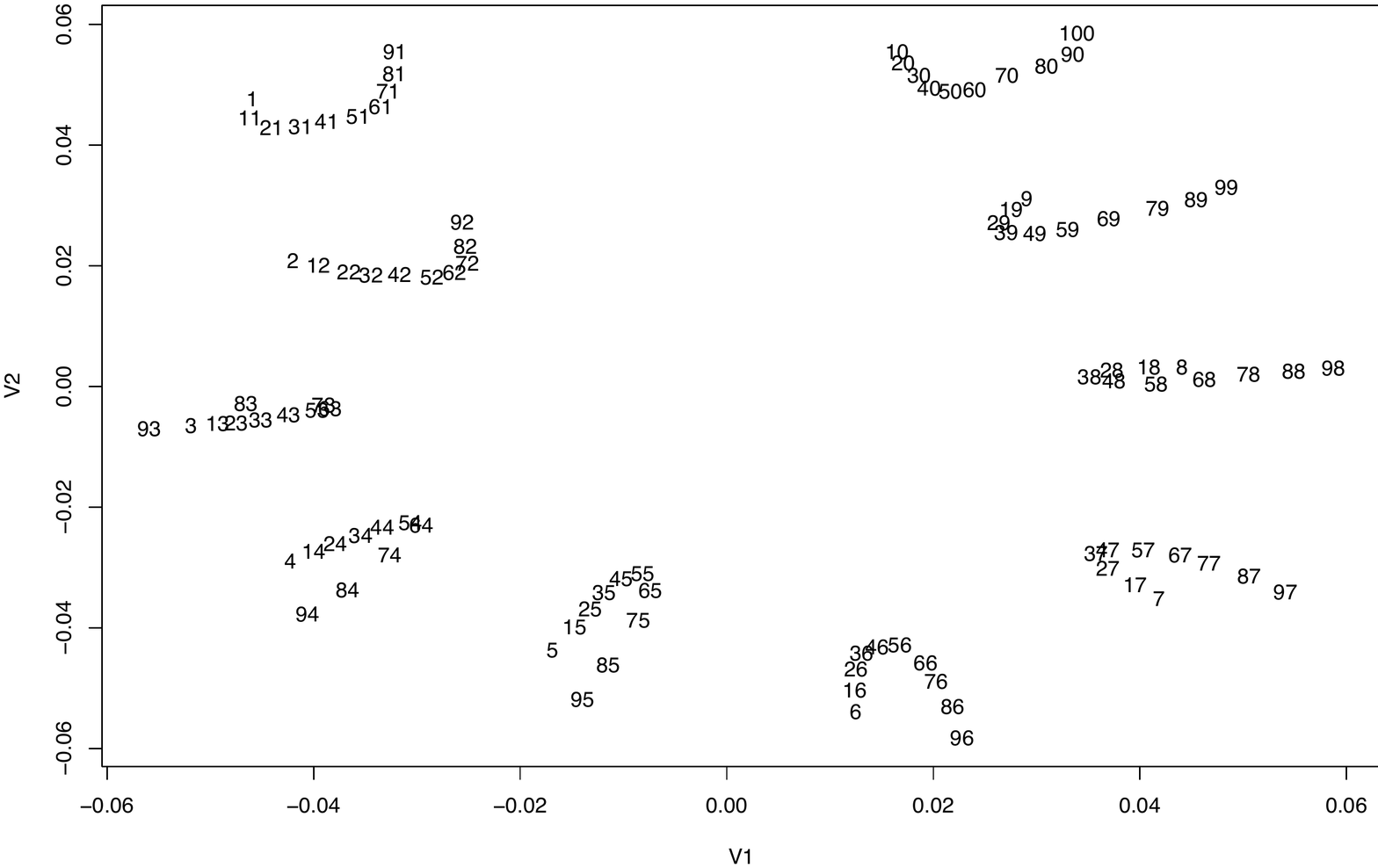} } & 
\subfigure{\includegraphics[width=0.4\linewidth]{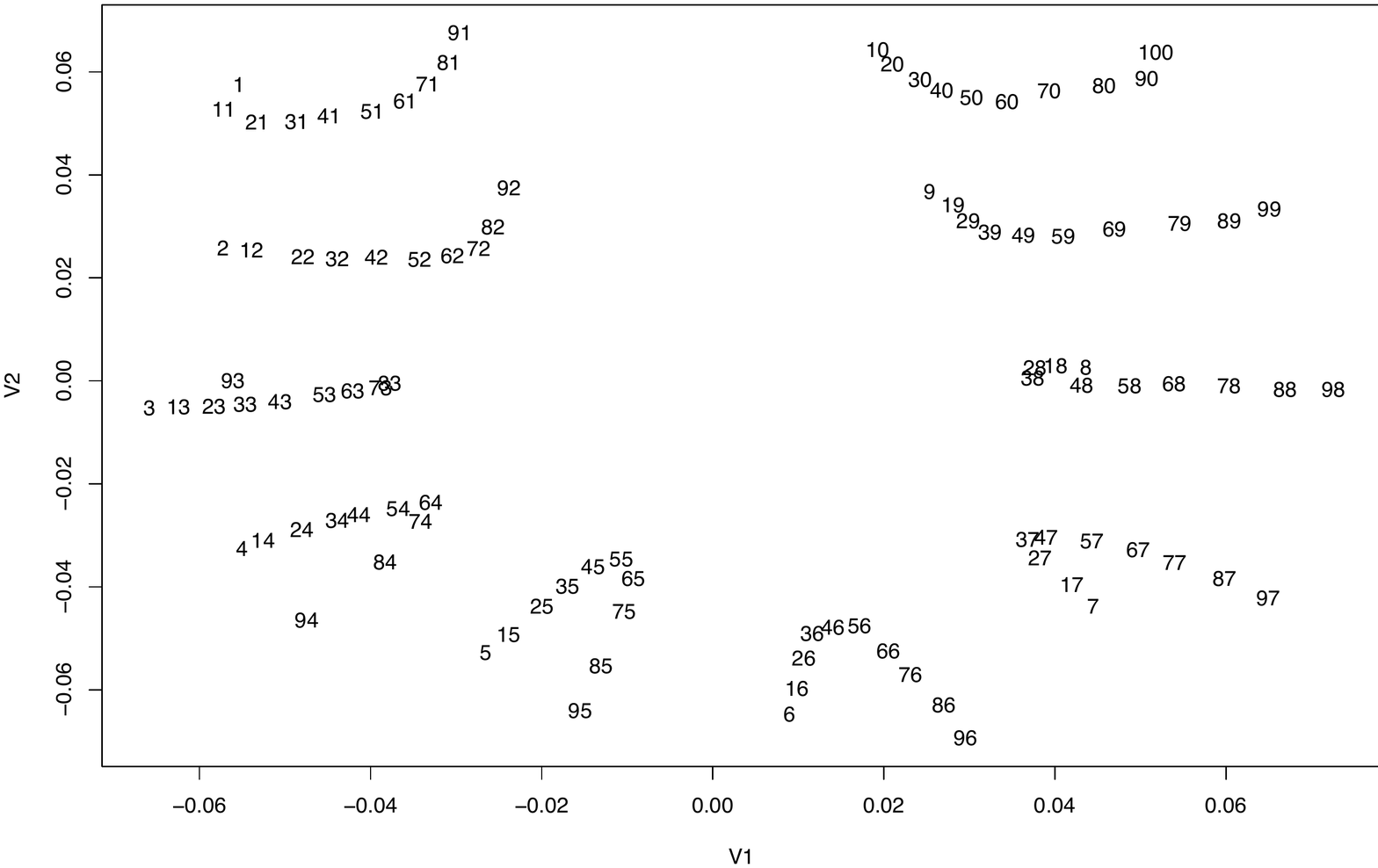} } \\
\subfigure{\includegraphics[width=0.4\linewidth]{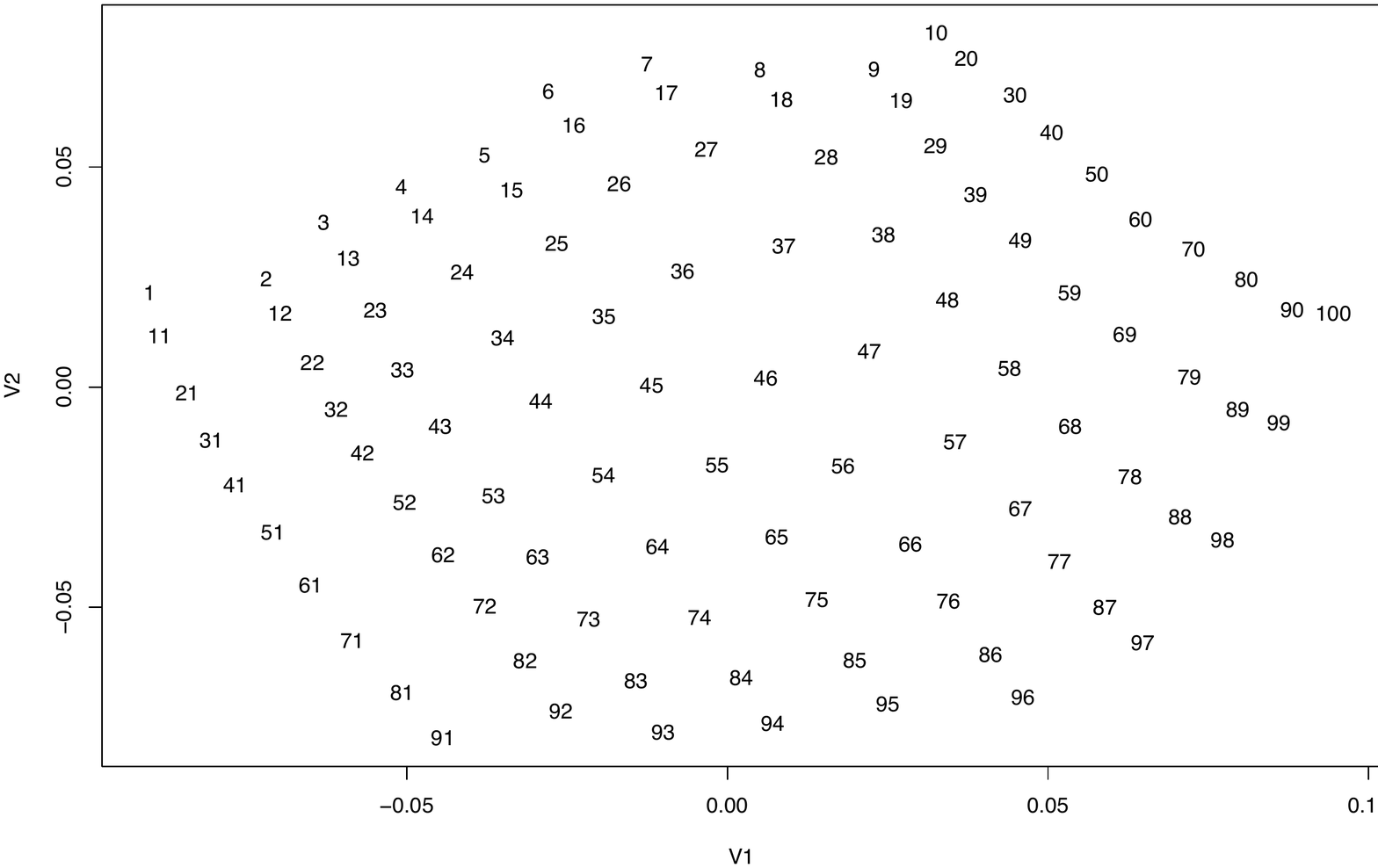} } &
\subfigure{\includegraphics[width=0.4\linewidth]{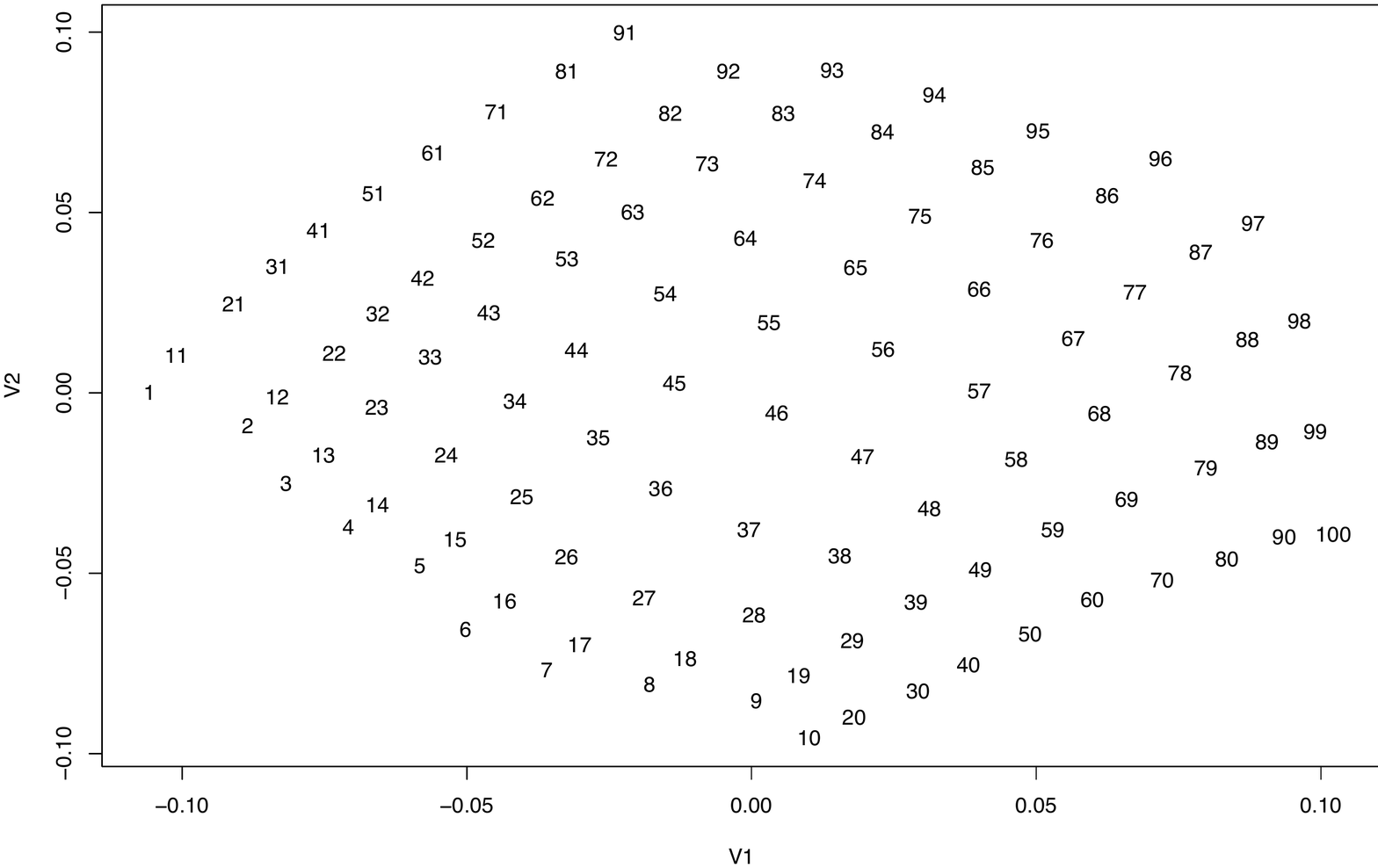} }
\end{tabular}\caption{Metric projections of informational distances between visual sensors.  Frames range from 1800 in the top left picture to 3000 in the bottom left picture. As time passes the sensor layout is optimized , from ~\cite{Olssonetal:2004:InformationTrade-Offs}.The two axis represents the informational distances in the two more statistically significant dimensions. As the agent moves from a simple environment, like the one represented by the horizontal lines, 
to a more complex one, like that represented by the rocks, the information distances between assume the grid structure in the picture. This corresponds to different physical layouts in different environments.(Courtesy of the authors).}
  \label{Fig:Layoutevo}
\end{figure}

\vspace{0.5cm}
\begin{center}
[Insert Figure~\ref{Fig:Layoutevo}]
\end{center}
\vspace{0.5cm}
It is worth noticing that this is just one example of the many performance indeces we may devise by weighting information metrics to balance 'exploitation' and 'exploration'.
Figure~\ref{Fig:Layoutevo}shows how the sensor layout approximate the distribution of single eyes in a composite insect eye: an hint in favor of 'strong' IDSO. The evolved sensory layouts seem realistic. This suggests, as in the Snakebot and similar examples, that we  have good reasons to believe something similar is at work in nature. The main disadvantage, again, of this method is that it is computationally demanding. This example illustrates how 'morphology matters' as not all sensor layouts are equally fit, and eventually an 'insect eye' layout emerges. Howewer, a limitation of the case shown, is that it does not inherently take the morphology of the agent into account, because there is only a flat 10 by 10 square and as a consequence important aspects of the body morphology and kinematics are not considered.

\begin{framed}
\begin{center}Different metrics related to Shannon Entropy\end{center}
Because distinct researchers use different flavours of informational measures derived from Shannon entropy, we summarize the ones we have quoted in this article:

\begin{enumerate}

\item {\bf Information flow or Transfer entropy.} 
Mutual information has some limitations when applied to time series analysis. As it is symmetric it does not allow to ascertain if X influences Y or the opposite. In other words it does not indicate the  direction of the infomation flows.
'Transfer entropy',~\cite{Schreiber:2000:Measuringinformation}, also known as 'information flow' circumvents this shortcoming.
Transfer Entropy is defined as:
\begin{equation}
\begin{array}{l}
TE = {h_2} - {h_1} =  - \sum\limits_{{x_{n + 1}},{x_n},{y_n}} {p\left( {{x_{n + 1}},{x_n},{y_n}} \right)} \log \left( {{x_{n + 1}}\left| {{x_n}} \right.} \right)\\
 + \sum\limits_{{x_{n + 1}},{x_n},{y_n}} {p\left( {{x_{n + 1}},{x_n},{y_n}} \right)} \log p\left( {{x_{n + 1}}\left| {{x_n},{y_n}} \right.} \right) = \\
\sum\limits_{{x_{n + 1}},{x_n},{y_n}} {p\left( {{x_{n + 1}},{x_n},{y_n}} \right)} \log p\left( {\frac{{{x_{n + 1}}\left| {{x_n},{y_n}} \right.}}{{{x_{n + 1}}\left| {{x_n}} \right.}}} \right)
\end{array}
\label{eq:xdiffMetTE}\end{equation}


The quantity $ {h_1} $ represents the entropy rate for the two systems, while $ {h_2} $ represents the entropy rate assuming that $ x  $ is independent from $ y $.
We define as entropy rate the amount of additional information needed to represent the next observation of one of the two systems.

\item {\bf Granger Causality.} 
The 'Granger causality', abbreviated sometimes as 'G-causality', is a form of 'causality' based on statistical tests, see ~\cite{Granger:1969:Investigating}.
A stochastic variable X, or more specifically the time series of its sampled values is said to 'Granger-cause' the stochastic variable Y, if the values in the time series of X influence the predictability of the future values of Y.
It has been found by Barnett, ~\cite{Barnettetal:2009:Grangercausalityandtransfer}, that Transfer entropy and Granger causality are equivalent for Gaussian processes.

\item {\bf Excess entropy or Predictive information.} 
In general Predictive information represents the possibility to predict a future value of a time series when we know a series of past values. In a completely random time series this quantity is zero.
For Markov processes it is given by: 
\begin{equation}
I\left( {{X_{t + \tau }};{X_t}} \right) = \left\langle {\log \frac{{p\left( {{x_{t + \tau }},{x_t}} \right)}}{{p\left( {{x_{t + \tau }}} \right)p\left( {{x_t}} \right)}}} \right\rangle  = \left\langle {\log \frac{{p\left( {{x_{t + \tau }}\left| {{x_t}} \right.} \right)}}{{p\left( {{x_{t + \tau }}} \right)}}} \right\rangle 
\label{eq:diffMetPI}\end{equation}

In other words, for Markov processes, the 'predictive information' of a time series is equal to the mutual information between the current and the next measured values.
This concept was actually proposed before by Crutchfield in~\cite{Crutchfield:1989:Inferring}, with the name of 'excess entropy'.

\end{enumerate}

\end{framed}

\section{How to deal with embodiment}	 

Genetic Algorithms, as population based non gradient search methods, can be seen as search methods for high dimensional spaces.  The performance index maximized in ~\cite{Olssonetal:2004:InformationTrade-Offs}, and the reinforcement learning method used in  ~\cite{Prokopenkoetal:2006:EvolvingSpatiotemporal}, are based on the brute force computation of informational metrics on the sensor and the actuators values. These metrics are in practice computed as binned summations of time series (i.e. you identify 'bins', a small number of consecutive time samples, and you compute the density of values representing the probabilities that are usesìd in the calculations) .
The result of the combination of an effective, but still computationally heavy algorithm like the Microbial GA  or in general any other optimization strategy, with a fitness function based on Shannon-like Informational Measures entail remarkable computational burdens.  This has limited so far the applications to toy-systems like those discussed above and in ~\cite{Olssonetal:2004:InformationTrade-Offs,Deretal:2006:Letitroll,Prokopenkoetal:2006:EvolvingSpatiotemporal}.
Moreover, 'pure' informational based metrics seem unsuitable for natural learning processes in natural intelligent agents for the same reasons.
It is likely that the search process is made simpler in nature, because part of the information processing is offloaded to the 'body dynamics'.
How can this be explained and represented in a formal way suitable to be exploited in an algorithm?
The main idea forwarded here is that the body shapes the computing essentially in two ways : 
\begin{enumerate}
\item it reduces the available phase space to a well defined subset of the possible movements in SE(3)
\item it exploits the symmetries in the possible motions
\end{enumerate}

\begin{figure}[!ht]
   \centering
      \includegraphics[width=\linewidth]{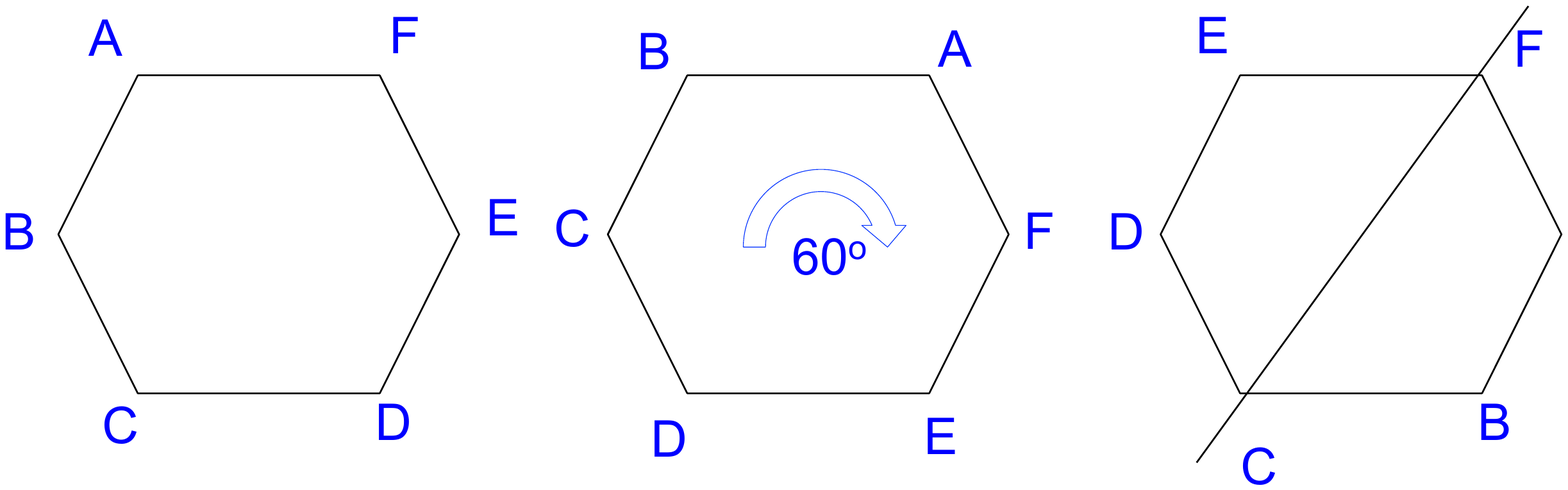}
   \caption{Examples of Symmetry Groups.The figure does not change if we rotate the exagon by 60 degrees or if we reflect the figure around the axis CF. Both this operations identify a symmetry group.}
  \label{Fig:Exofsymmetry}
\end{figure}

\begin{figure}[!ht]
   \centering
      \includegraphics[width=\linewidth]{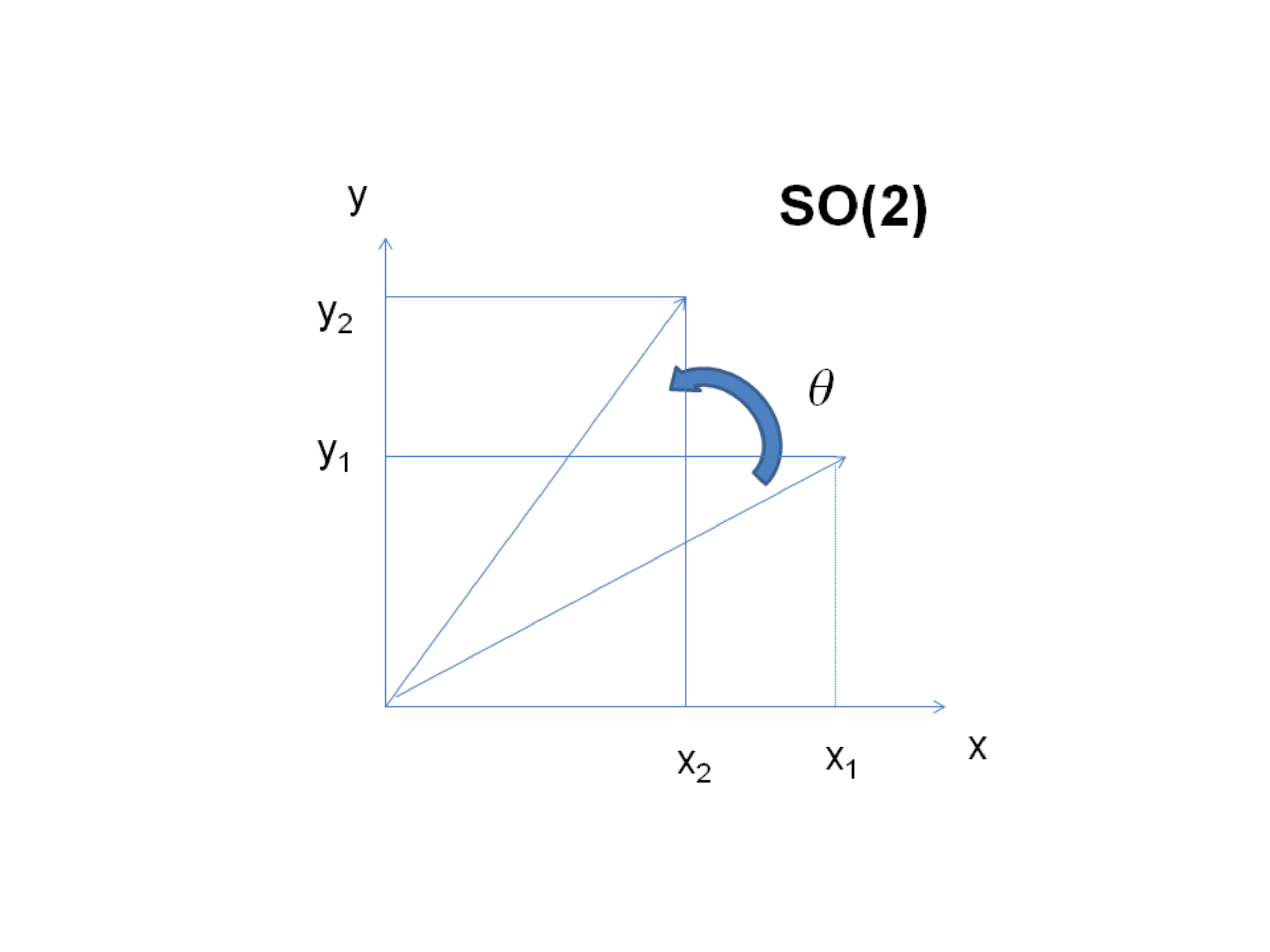}
   \caption{The group of continuous rotations in the plane. The group 'operation' is given by the counter clockwise rotation of a point $ P_1 (x_1, y_1)  $ around of the origin of the coordinates by an angle $ \theta $ to the new position $ P_2 (x_2, y_2)  $. The new group operation can be rapresented by equation~\eqref{eq:xdef12b}. }
  \label{Fig:SO2}
\end{figure}

\begin{framed}
\begin{center}Symmetries, Lie Groups and SE(3)\end{center}
A symmetry in a system behavior (in particular in its motions) expresses the fact that the system doesn't change when subject to (certain) changes,  see for example Figure~\ref{Fig:Exofsymmetry}.
Symmetries can be expressed mathematically with the concept of a 'Group'.
A Group 'G' is a set of objects that can be combined by a binary operation (called 'group multiplication' or composition rule, denoted by  '$ \circ $'). 

\vspace{0.5cm}
\begin{center}
[Insert Figure~\ref{Fig:Exofsymmetry}]
\end{center}
\vspace{0.5cm}



'Elements' of the group are the objects that form the group (generally denoted by 'g')
A 'Generator': is a (minimal) subset of elements that can be used to obtain 	(by means of group 'multiplication') all the elements of the group)
More precisely, a group 'G' is a set that:
\begin{enumerate}

\item{} is closed under multiplication ($ \ \circ \ $) - if  a,b are in G then $ \ a \circ b\ $  is also in G 
\item{} contains an identity element 'I'
\item{} the inverse of each element is also part of the group ($ \ g \circ {g^{ - 1}} = I \ $ )
\item{} the group composition rule, the 'multiplication' is associative, i.e.  $ \ a \circ (b \circ c) = \left( {a \circ b} \right) \circ c \ $  (but not necessarily commutative).
\end{enumerate}
Figure~\ref{Fig:Exofsymmetry} shows a discrete group in which elements can be counted (i.e. they have an integer number of elements).
More relevant to our problem are continuous groups. In a continuous group the Elements are generated by continuously varying a number (one or more) of parameters.
Combinations of rotations and traslations in space, or on a plane can be represented continuous groups (actually they are instances of Lie groups).
As in the following example of a simple Lie Group, see Figure~\ref{Fig:SO2},  formally named as the 'Group of all Rotations in 2D space - SO(2) group' and essentially representing a rotation of a (rigid) body on a plane.

\vspace{0.5cm}
\begin{center}
[Insert Figure~\ref{Fig:SO2}]
\end{center}
\vspace{0.5cm}


\begin{equation}\nonumber\\
 \left[ {\begin{array}{*{20}{c}}
{x2}\\
{y2}
\end{array}} \right] = \left[ {\begin{array}{*{20}{c}}
{\cos \theta }&{ - \sin \theta }\\
{\sin \theta }&{\cos \theta }
\end{array}} \right]\left[ {\begin{array}{*{20}{c}}
{x1}\\
{y1}
\end{array}} \right]
\label
{eq:xdef12a}\end{equation}
\begin{equation}
U(\theta ) = \left[ {\begin{array}{*{20}{c}}
{\cos \theta }&{ - \sin \theta }\\
{\sin \theta }&{\cos \theta }
\end{array}} \right]
\label
{eq:xdef12b}\end{equation}

More formally a Lie Group is defined as a group whose elements can be parameterized by a finite number of parameters i.e. it is a continuous group that satisfies the two equivalent properties:
\begin{enumerate}

\item{} if $ \ g\left( {{a_i}} \right) \circ g\left( {{b_i}} \right) = g\left( {{c_i}} \right)\ $  then  $ \ {c_i} \ $ is an analytical function of $ \ {a_i}\ $  and $ \ {b_i}\ $  
\item{}the group manifold is differentiable
\end{enumerate}
For a given multi rigid body structure, such as an arm, a leg or even a hand, the number of possible motions is limited to the composition of a comparatively small number of motions, represented by Lie (sub)groups. This fact greatly reduces the part of the system's phase space that has to be searched. As we will see in the next section, and in Appendix A and B, in certain assumptions, it is also possible to approximate the informational metrics by means of closed form expressions.
The ensemble of all the possible motions of a body in the usual tridimensional space (in general compositions of rotations and translations) is called SE(3). SE(2) is the  ensemble of all the possible motions of a body on a plane (a bidimensional space).
\end{framed}

\subsection{Main Concepts}
\begin{figure}[!ht]
\centering
\mbox{\subfigure{\includegraphics[width=0.4\linewidth]{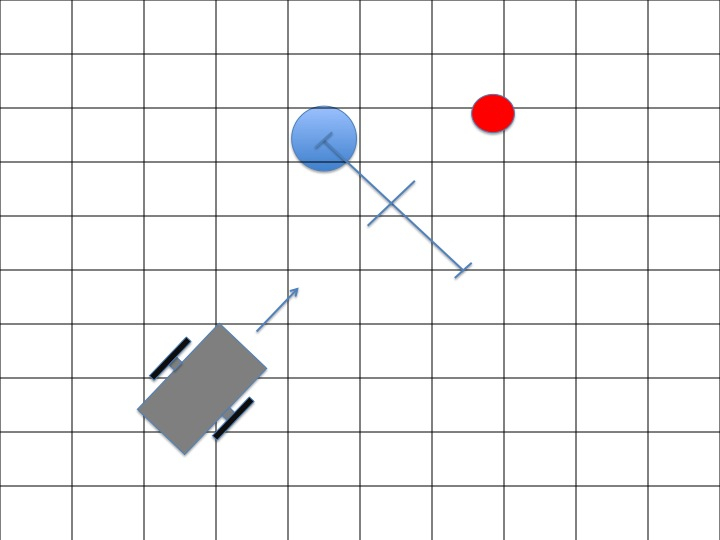}}\quad
\subfigure{\includegraphics[width=0.4\linewidth] {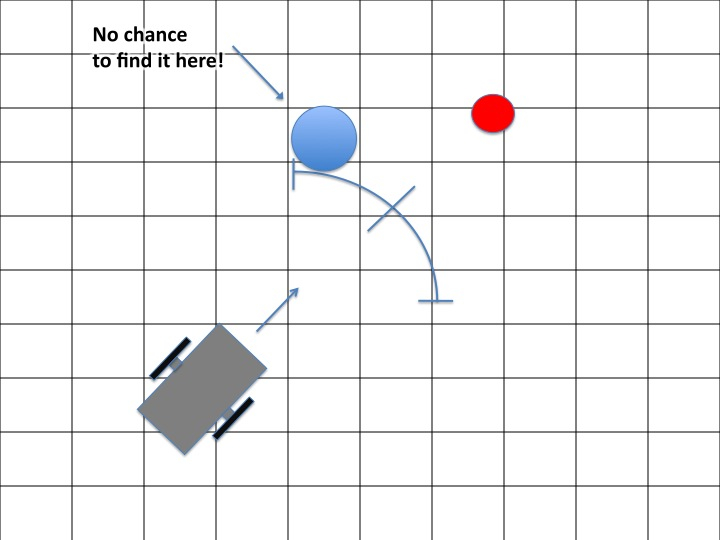}}}
\caption{Statistical distribution on Lie groups. If we have a Gaussian multivariate distribution on a Lie group, the 'projection', marginalization, in the usual space is not a Gaussian, but a 'Banana' distribution, as we can see on the right side of the figure. In practice this means that, if we have a differential wheeled mobile robot moving towards a target, like in this picture, and we estimate its position at a given time, schematized by the line crossing the arrow, we will estimate, if we assume a Gaussian distribution of position errors (picture on the left), a non zero probability to find it in a place, where, if we consider the group structure of motion of a physical body (picture on the right), has practically zero probability to be found.} \label{Fig:Liegaussianvsbanana}
\end{figure}

\begin{figure}[!ht]
\centering
\mbox{\subfigure{\includegraphics[width=0.4\linewidth]{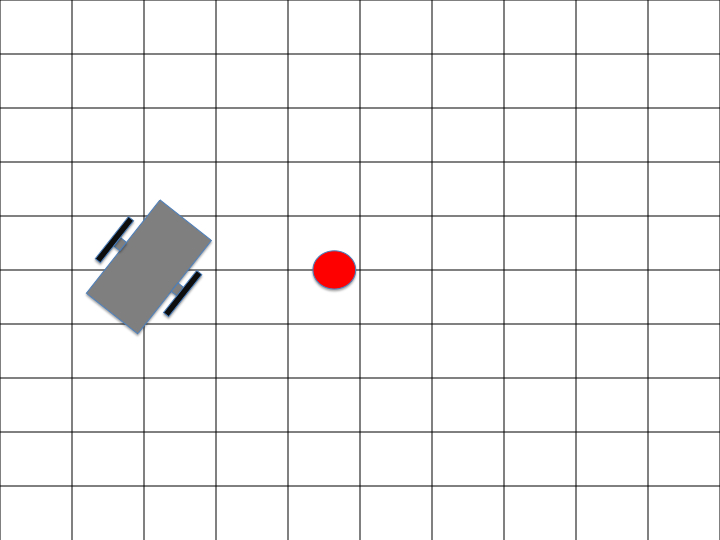}}\quad
\subfigure{\includegraphics[width=0.4\linewidth] {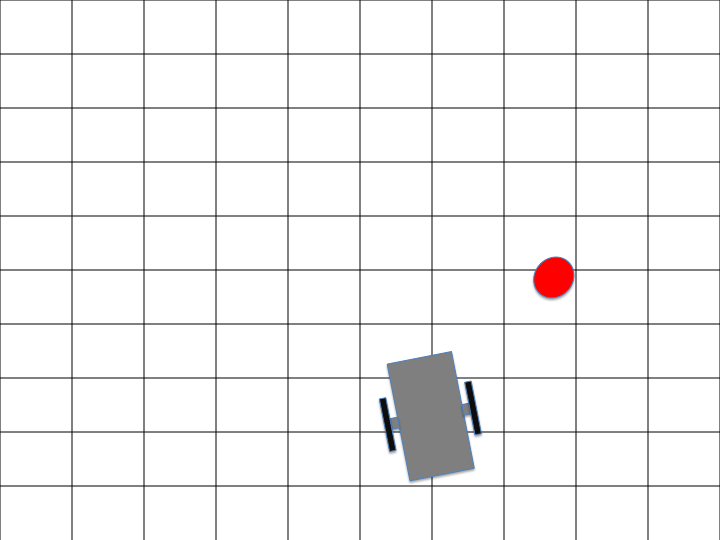}}}
\caption{SE(2) simmetry. The two situations in this figure are identical, for example when we calculate Predictive information for this mobile robot, we do not need to calculate it twice (and for all the possible positions and orientations in SE(2)). A similar situation occurs in SE(3). These fact are captured by Lie groups.} \label{Fig:Liesymmetry}
\end{figure}

The main contribution of this paper is the observation that it is possible to design information driven self organizing control processes exploiting, through the Lie group formalism, the embodiment of the controllers into a physical macroscopic body. This is on the one hand more correct in the representation of the uncertainty connected to the embodied agent,see ~\cite{LongChirikjian:2012:TheBanana}, on the other hand more computationally effective.
It is more correct, because if we apply the central limit theorem to a physical macroscopic body we have a Gaussian in SE(3), the 'g's and if we marginalize in x,y,z we have a distribution resembling a banana not a mexican hat (see Figure~\ref{Fig:Liegaussianvsbanana} ). It is more effective from the computationel standpoint because by exploiting the symmetries in the motion we can in many cases dramatically reduce the computational burden of computing the informational metrics we want to optimize: in Figure~\ref{Fig:Liesymmetry}, the two situations are exactly equivalent, a fact that is captured by the Lie group formalism.
This methodology is made possible by merging mathematical results coming from two different lines of research, the IDSO line of research and the stochastic kinematics line of research.
In summary:
\begin{enumerate}

\item 'Banana' not Gaussian distributions
\item The 'Bananas' can be computed with (relatively) limited effort
\item The optimizations can be performed by a population based
evolutionary programming scheme (among the many possible choices)

\end{enumerate}

In the next paragraphs,we  first revisit the example introduced in the previous section in light of these concepts and then we show how the transfer entropies in ~\cite{LungarellaSporns:2006:MappingInformationFlow} can be computed more efficiently.

\vspace{0.5cm}
\begin{center}
[Insert Figure~\ref{Fig:Liegaussianvsbanana}]
\end{center}
\vspace{0.5cm}

\vspace{0.5cm}
\begin{center}
[Insert Figure~\ref{Fig:Liesymmetry}]
\end{center}
\vspace{0.5cm}

\subsection{Revisiting the evolution of sensory layout}

In this section we describe an embodied version of the model of evolution of sensory layouts proposed in  ~\cite{Olssonetal:2004:InformationTrade-Offs}.
The possible motions of a physical body are structured in terms of Lie groups:

\begin{enumerate}
\item {\bf articulated rigid multi body systems} they are constrained to a finite group of roto traslations mathematically expressed by a finite number of Lie groups, subgroups of the general Lie group. 
\item {\bf deformable systems}  we focus on infinitesimal motion the possible motion of the material particle of a deformable continuous  body are still constrained to roto traslations. Information metrics are computed on Lie groups with Lie algebra instead of doing that on 'flat' $ \mathbb{R}^n $  spaces (the configuration space of a physical system can actually be regarded as a curved manifold embedded in $ \mathbb{R}^n $). 
\end{enumerate}
The approach in  ~\cite{Olssonetal:2004:InformationTrade-Offs} , or similar ones, can be modified when the algorithm is 'made aware' of the body morphology. This can be achieved as follows. The evolutionary algorithm optimizes the performance index ic(S+C), a weighted sum of two metrics: one representing redundancy (through mutual information), and one representing the predictive power of the sensory-motor control system. S and C  are the stochastic vector variables representing the state of the sensors and the controller (including the actuators), see ~\cite{TouchetteLloyd:2003:Informationtheoreticapproach}. 
There are three important differences in the approach proposed here, with respect to  
~\cite{Olssonetal:2004:InformationTrade-Offs} : 
\begin{enumerate}
\item we deal with emergent controllers embedded in a co-evolving physical body structure (not only sensory layouts) 
\item we analyze how the body shapes the controllers by exploiting the powerful Lie group formalism and related concepts 
\item as we deal with controllers we refer to a more suitable information metric: the predictive information.
\end{enumerate}
 
This approach takes care of the limitations of the body dynamics, by considering the kinematic structure to which the system variables comply. 

We may define a weighted performance index, weighing the predictive power of the overall system through the predictive information PI (a metric related to Shannon entropy) and the redundancy through mutual information between the different sensors:
 
\begin{equation}
ic(S + C) = \sum\limits_{X \in \left( {S + C} \right)} {\sum\limits_{Y \in \left( {S + C} \right)} {\left( {{w_{mi}}I\left( {X;Y} \right) + {w_{pi}}PI} \right)} } 
\label{eq:xdef20}\end{equation}

\begin{equation}
I\left( {X;Y} \right) = H\left( X \right) - H\left( {X\left| Y \right.} \right) = H\left( Y \right) - H\left( {Y\left| X \right.} \right)
\label{eq:xdef21}\end{equation}

\begin{figure}[!ht]
   \centering
      \includegraphics[width=\linewidth]{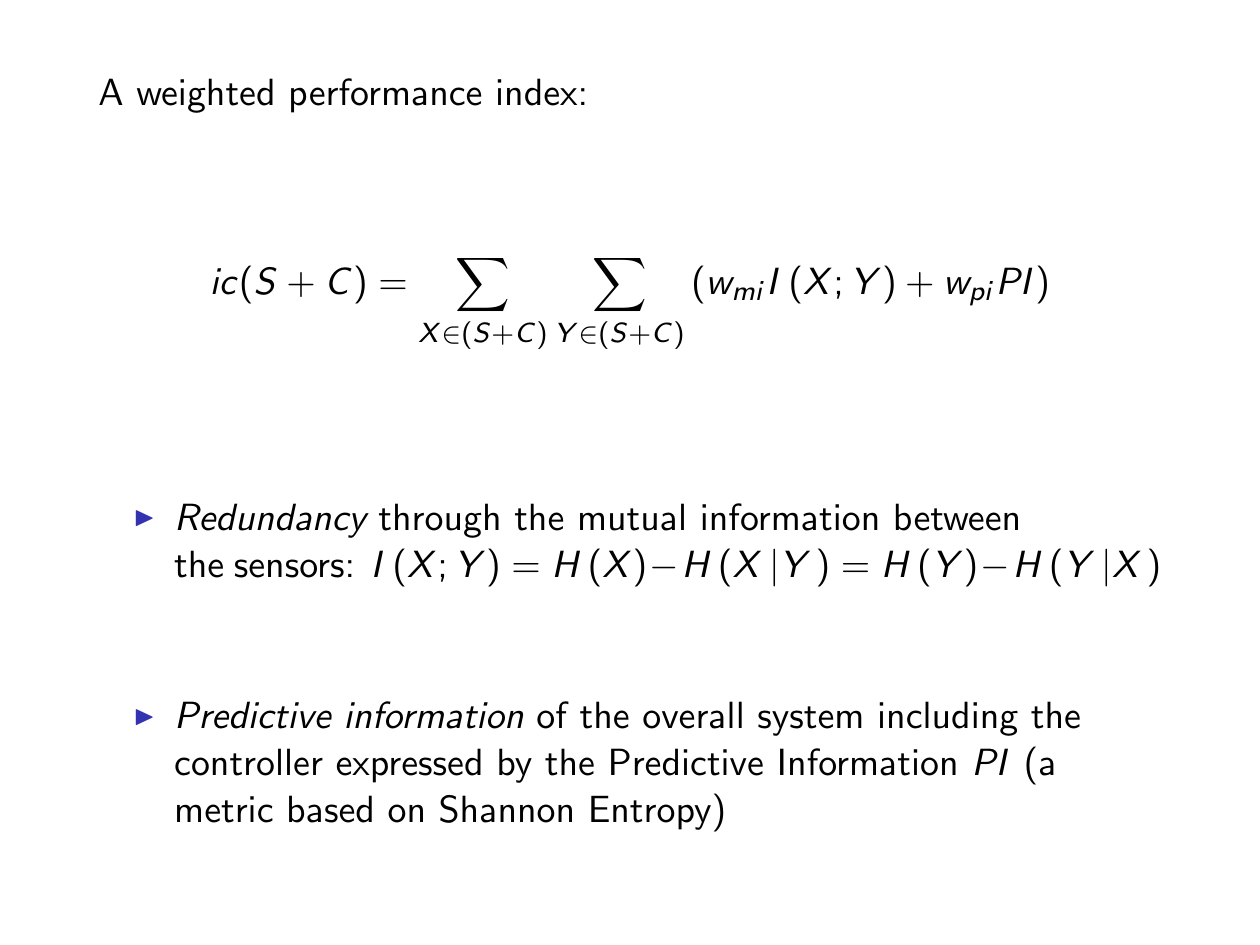}
   \caption{The approach considering the embodiment.The performance index $ ic(S+C) $ used to guide the evolutionary algorithm is a function of the stochastic vector  variables $ S $ representing the temporal series of the sensor values and $ C $ representing the temporal series of the vector values of the internal state of the controller and of actuation values, see~\cite{TouchetteLloyd:2003:Informationtheoreticapproach} and~\cite{Bonsignorio:2007:Preliminary} for the details. Following the example in~\cite{Olssonetal:2004:InformationTrade-Offs} we maximize a weighted sum of a metric representing 'redundancy' in the system with a metric measuring the 'diversity'. The second one is given by 'Predictive information' as we see sensing and actuation together. Both metrics are calculated considering the Lie group structure of motion by means of the formulas derived in the appendixes and discussed in the paper. }
  \label{Fig:Embodiedevo}
\end{figure}

Figure~\ref{Fig:Embodiedevo}  shows how we can implement the suggested procedure by exploiting the (Lie) symmetries of the kinematic structure.
\vspace{0.5cm}
\begin{center}
[Insert Figure~\ref{Fig:Embodiedevo}]
\end{center}
\vspace{0.5cm} 

The predictive information is given by equation \eqref{eq:xdef30a}, if we, for example, consider a kinematic serial chain made of a series of rigid bodies identified by a set of frames.
In general the motion structure provide the constraints so that the predictive information can be computed directly and in closed form. In addition, we can derive a closed form reinforcement learning rule maximizing it (with the assumption of 'tight' Gaussian distributions); or (in more general cases  without special assumptions) to simplify the computation. The predictive information concept, also known as excess entropy was introduced by Crutchfield and can be seen as a measure of complexity, ~\cite{Crutchfield:1989:Inferring}. Der uses the time loop error, as a measure of complexity,~\cite{Deretal:2006:Letitroll}, for time series.
The details can be found in Appendix B.
Recognizing that the mobility space of a physical structure is actually a subspace of the $ \mathbb{R}^n $  Cartesian space, has the potential, by reducing dramatically their computation cost, to make the computationally heavy IDSO methods applicable to non trivial physical structures coping with their greater limitations. The trade-off is that this subspace is actually a curved manifold and that the 'operations' on it are not commutative, yet a mature and powerful mathematical theory is available: Lie group theory.

\subsection{Revisiting the discussion in section 2}
The observations made above are true also when we want to analyze the informational trade-offs between controllers and body dynamics.
Let us consider again, for example, one of the physical instantiations of sensory-motor coordination shortly described in section 2, A3 in Figure~\ref{Fig:Expsettings}.\\
We have a differential wheel cart constrained to move on a plane, and then we have a pan and tilt camera mounted on it. Thus we have a SE(2) group symmetry and two rotational joint (corresponding to pan and tilt movements of the camera). We can then identify the motion group of the whole system as:
\begin{equation}
{G_{A3}} = SE(2) \times SO(2) \times SO(2)
\label{eq:xdefGB3}\end{equation}
If we assume a Gaussian distribution on the $ g \in {G_{A3}} $ the (differential) entropy will be given by equation \eqref{eq:xdef16}, the predictive information by equation \eqref{eq:xdef30a}.
\clearpage
\begin{framed}

\begin{center}Useful Lie group operators and properties of rigid body motion\end{center}
Exponential operators prove very useful in differential geometry, yet they are not widely known and used.
Here, we review their basic definitions and properties.
The Euclidean motion group SE(3) is the semidirect
product of R3 with the special orthogonal group SO(3) We
define an element 'g' of SE(3) as $ \ g = \left( {\bar a,A} \right) \in SE(3)\ $, where $ \ a \in {R^3}\ $
and $ \ A \in SO\left( 3 \right)\ $.
\\For any $ \ g = \left( {\bar a,A} \right)\ $ and $ \ h = \left( {\bar b,B} \right)\ $ the group
composition law is written as: 
\begin{equation}
g \circ h = \left( {\bar a + \bar b,AB} \right)\ \label
{eq:xLieUs1}\end{equation} while the
inverse of g is given by: \begin{equation} {g^{ - 1}} = \left( { - {A^T}\bar a,{A^T}} \right)
\label
{eq:xLieUs2}\end{equation}
An alternative representation is given by 4x4 homogenous matrices
of the form:
\begin{equation} H\left( g \right) = \left( {\begin{array}{*{20}{c}}
A&{\bar a}\\
{{0^T}}&1
\end{array}} \right)
\label
{eq:xLieUs3}\end{equation}

In this case the group
composition law is given by matrix multiplication.
For small translational/rotational displacements from 
the identity along (translational) / about (rotational) the $ {i^{th}} $ coordinate axis the
homogeneous transformation matrix is given approximately by:
\begin{equation} {H_i} \buildrel \wedge \over = \exp \left( {\varepsilon {{\tilde E}_i}} \right) \approx {I_{4X4}} + \varepsilon {\tilde E_i}
\label
{eq:xLieUs4}\end{equation}

where $ \ {I_{4x4}}\ $ is the identity matrix and:

\begin{equation}\nonumber\\
{\tilde E_1} = \left( {\begin{array}{*{20}{c}}
0&0&0&0\\
0&0&{ - 1}&0\\
0&1&0&0\\
0&0&0&0
\end{array}} \right)
~~\mbox{ }~~ {\tilde E_2} = \left( {\begin{array}{*{20}{c}}
0&0&1&0\\
0&0&0&0\\
{ - 1}&0&0&0\\
0&0&0&0
\end{array}} \right) \
\end{equation}

\begin{equation}\nonumber\\
{\tilde E_3} = \left( {\begin{array}{*{20}{c}}
0&{ - 1}&0&0\\
1&0&0&0\\
0&0&0&0\\
0&0&0&0
\end{array}} \right)
~~\mbox{ }~~
{\tilde E_4} = \left( {\begin{array}{*{20}{c}}
0&0&0&1\\
0&0&0&0\\
0&0&0&0\\
0&0&0&0
\end{array}} \right)\
\end{equation}

\begin{equation}
{\tilde E_5} = \left( {\begin{array}{*{20}{c}}
0&0&0&0\\
0&0&0&1\\
0&0&0&0\\
0&0&0&0
\end{array}} \right)
~~\mbox{ }~~
{\tilde E_6} = \left( {\begin{array}{*{20}{c}}
0&0&0&0\\
0&0&0&0\\
0&0&0&1\\
0&0&0&0
\end{array}} \right)\
\label
{eq:xLieUs5}\end{equation}

Large motion can be obtained by exponentiating these
matrices.
It is useful to describe elements of SE(3) with the
exponential parametrization:
\begin{equation}
g = g\left( {{x_1},{x_2},...{x_6}} \right) = \exp \left( {\sum\limits_{i = 1}^6 {{x_i}{{\tilde E}_i}} } \right)
\label
{eq:xLieUs6}\end{equation}

If we define the 'vee' operator such that:
\begin{equation}
{\left( {\sum\limits_{i = 1}^6 {{x_i}{{\tilde E}_i}} } \right)^V} =
\left( {{x_1},{x_2},{x_3},{x_4},{x_5},{x_6}} \right)^T
\label
{eq:xLieUs7}\end{equation}

as a consequence the total vector can be obtained
as:
\begin{equation} 
x = {\left( {\log g} \right)^V}
\label
{eq:xLieUs8}\end{equation}
\end{framed}

\section{Discussion and future work}
	
\begin{figure}[!ht]
   \centering
      \includegraphics[width=0.75\linewidth]{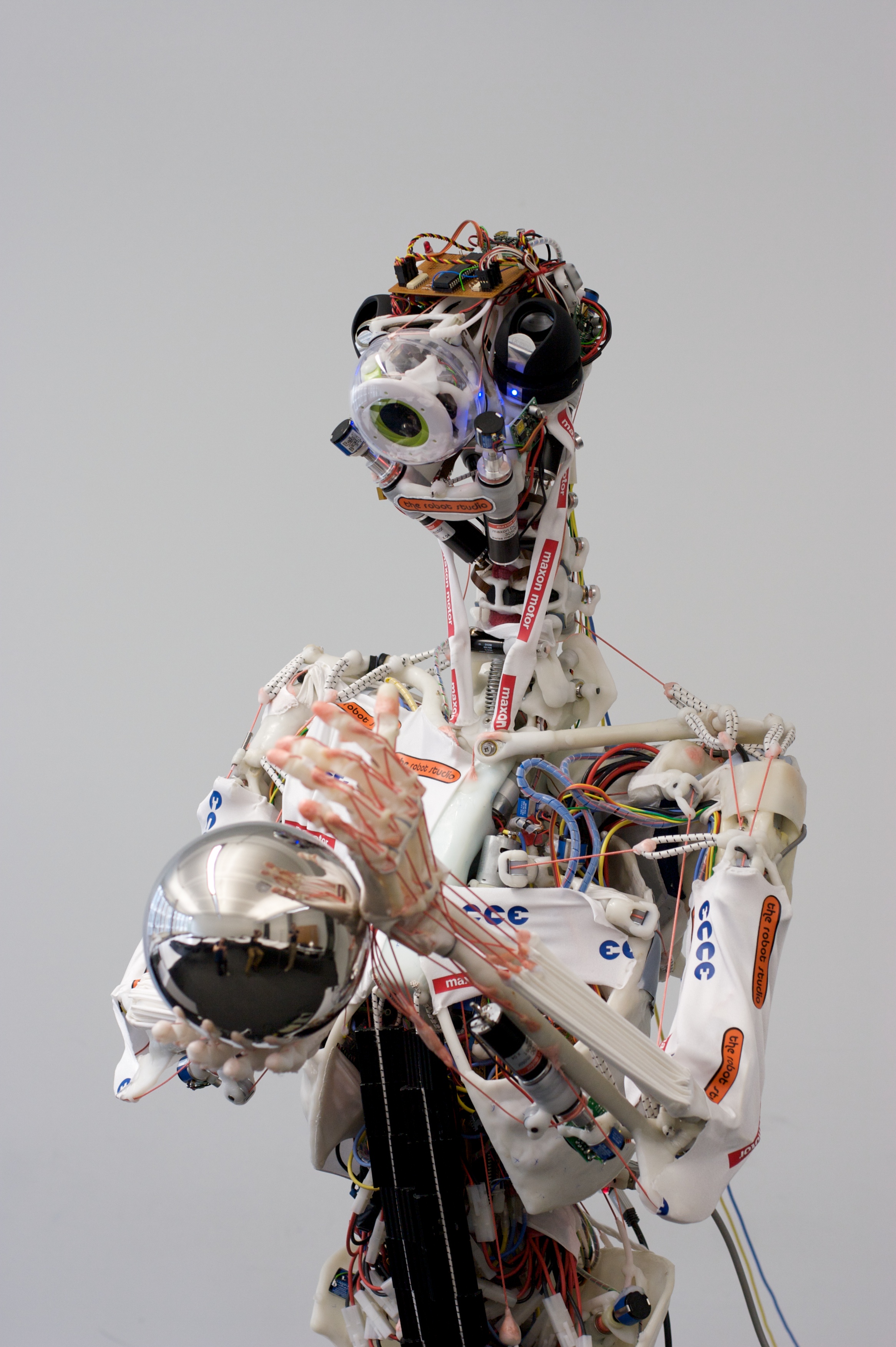}
   \caption{The 'Ecce' robot.The Ecce robot is a humanoid with a human like tendon driven actuation of arms and hands and a deliberately imprecise body structure.}
  \label{Fig:Ecce}
\end{figure}

The examples proposed above can be more easily managed if we assume concentrated Gaussian distributions for the state variables and rigid multi body dynamics. While the first assumption may not be particularly limiting in the context of controlled physical systems, the latter is not directly applicable to deformable 'soft', natural or artificial, cognitive systems. In our perspective to consider a deformable distributed controlled body as a rigid multi body one, might be seen as a way of freezing degrees of freedom (d.o.f.), which has been proposed as a viable control approach for soft robots. 
In other words: an emerging controller like those described here, can be applied to embodied intelligent agents which can be approximated by a rigid multi body structure with concentrated uncertainties and elasticities, like Snakebot, Figure~\ref{Fig:Snakebot}, or Ecce, Figure~\ref{Fig:Ecce}; a fully distributed continuously deformable structure like that of an octopus will require more work, in particular from the mathematical  standpoint.
\vspace{0.5cm}
\begin{center}
[Insert Figure~\ref{Fig:Ecce}]
\end{center}
\vspace{0.5cm}
 
It should be noted that, while in\cite{Olssonetal:2004:InformationTrade-Offs}, the outcome of the evolutionary algorithm is a 'fit' sensor layout, in systems like that represented in Figure~\ref{Fig:Ecce}, the outcome will be a 'fit' configuration of the relative distances and orientations of, for example, eye and arm or hand, between themselves and from the sensors and the objects in the environment.
Are these methods or analogous ones exploited in nature? Is this the reason why the primate brain uses sometimes affine or quasi affine geometries,~\cite{MaozFlash:2009:Complex}, for motion planning?
 

\subsection{Interpreting the 'freezing' mechanisims}

It is reasonable to think that the d.o.f. freezing mechanism might also be better understood and detailed from this perspective.
For example, a wheeled mobile robot with a pan and tilt camera like A3 in Figure~\ref{Fig:Expsettings}, by freezing the pan and tilt motion of the camera, reduces itself to SE(2), a plane motion, the highly symmetrical mobility space  depicted in Figure~\ref{Fig:Liesymmetry} and then it can more economically compute its moves.
Actually the local properties of a deformable structure at a given time can be approximated infinitesimally by a rigid multi body structure; from this perspective the discussion above still holds, if we consider the physical approximating structure changing in time and if we see this approximating rigid body shape as part of the optimization process.
In other terms to locally decrease the stiffness of parts of a human arm during unprecise motion, for example in grasp pre-shape, might be seen as a way to transform a comparatively loose structure, with a high associated entropy, into a rigid multi body one (a rag-doll in computer graphocs jargon), with a remarkably lower associate entropy, the computational burden.
On the contrary during precise grasping the additional degrees of freedom provided by the deformability of the fingers make more likely a proper grasp.

\subsection{A More General Formalism}

These methods, when fully developed, may have potentially disruptive applications ranging from under-actuated locomotion to soft visual grasping systems, by the online optimization of the full sensing and actuation loop, based on embodied information driven self organization processes and exploiting the body deformability to ease optimization.
The generalization on this approach to fully deformable structures  is part of future work and involves the application of a 'cleaner' mathematical setting coming from differential geometry, and also from the theory of fiber bundles and connections, ~\cite{Olver:2000:ApplicationsofLieGroup,Kolaretal:1994:NaturalOperations}. 

\subsection{A frame of reference problem?}

Lie groups characterize the stochastic kinematics (and dynamics) of physical bodies and we suggest that they are incorporated in emerging controller schemes for sensory-motor coordination. We see in the simplification coming from the adoption of Lie groups modeling approach a palpable and measurable benefit of the incorporation of body morphology to ease computation. In other words we regard it as a quite general and quantifiable example of morphological computation.
As a consequence, as Lie groups are useful for a  basic representation of the sensory-motor coupling of an embodied intelligent agent with the environment, an important aspect of the agent learning will be the identification of the 'structure of the space' or rather the representation of what an agent can do with its individual body, its 'bodily affordances', see for example the work from O'Regan, ~\cite{PhiliponaRegan:2004:Perception}, showing the emergence of the 'space awareness', the abstract space 'representation', is a consequence of bodily affordances in an embodied agent.



\section{Conclusions}

In this paper, we showed a method to develop quantifiable information driven, self organizing sensory-motor coordination processes. This method may allow  to shape the emergence of control on the basis of the body morphology. This is possible at the price of a comparative abstraction and will require more work to be fully applicable to continuously deformable structures as exemplified by octopi.  It seems that something like what we propose here might be implemented to guide the 'freezing' mechanism of a fully or partially deformable, natural or artificial, sensorial and actuation system.

A Lie group representation is a way to represent the 'body affordances', the body morphology and the morphological computation,  to the 'brain' of the artificial agent, without doing a real distinction between information processing and the dynamics.

\section*{APPENDIX A. Stochastic Kinematics of Rigid Bodies}

We review here some known relations from the stochastic kinematics of rigid bodies. 
If we consider a vector function $ f\left( {\bar x} \right) $ with 
$ \bar x \in \mathbb{R}^n $ we can define $ \mu $, as in \eqref{eq:xdef131}, and $ \Sigma $, as in \eqref{eq:xdef132}: 
 
\begin{equation}	
0 = \int\limits_{{R^n}} {\left( {\bar x - \mu } \right)} f\left( {\bar x} \right)d\bar x
 \label{eq:xdef131}\end{equation}
 
\begin{equation} 
\Sigma  = \int\limits_{{R^n}} {\left( {\bar x - \mu } \right){{\left( {\bar x - \mu } \right)}^T}f\left( {\bar x} \right)} d\bar x
\label{eq:xdef132}\end{equation}

We have the multivariable Gaussian distribution:

\begin{equation}
f\left( {\bar x;\mu ,\Sigma } \right) = \frac{1}{{c\left( \Sigma  \right)}}\exp \left[ { - \frac{1}{2}{{\left( {\bar x - \mu } \right)}^T}{\Sigma ^{ - 1}}\left( {\bar x - \mu } \right)} \right]
\label{eq:xdef133}\end{equation}

where:

\begin{equation} 
c\left( \Sigma  \right) = {\left( {2\pi } \right)^{{\raise0.7ex\hbox{$n$} \!\mathord{\left/
 {\vphantom {n 2}}\right.\kern-\nulldelimiterspace}
\!\lower0.7ex\hbox{$2$}}}}{\left| {\det \Sigma } \right|^{{\raise0.7ex\hbox{$1$} \!\mathord{\left/
 {\vphantom {1 2}}\right.\kern-\nulldelimiterspace}
\!\lower0.7ex\hbox{$2$}}}}
\label{eq:xdef134}\end{equation}

In a similar way we can define a function $ f(g) $ with $ g \in G $, and two quantities $ \mu $, defined as in \eqref{eq:xdef135}, and $ \Sigma $, as in \eqref{eq:xdef136}: 

\begin{equation}
\int\limits_G {{{\log }^V}} \left( {{\mu ^{ - 1}} \circ g} \right)f\left( g \right)dg = 0
\label{eq:xdef135}\end{equation} 

\begin{equation} 
\Sigma  = \int\limits_G {{{\log }^V}\left( {{\mu ^{ - 1}} \circ g} \right)} {\left[ {{{\log }^V}\left( {{\mu ^{ - 1}} \circ g} \right)} \right]^T}f(g)dg
\label{eq:xdef136}\end{equation}

We have the multivariable Gaussian distribution on $ G $:

\begin{equation}
f\left( {g;\mu ,\Sigma } \right) = \frac{1}{{c\left( \Sigma  \right)}}\exp \left[ { - \frac{1}{2}{{\bar y}^T}{\Sigma ^{ - 1}}y} \right]
\label{eq:xdef137}\end{equation}

where:

\begin{equation}
y = \log {\left( {{\mu ^{ - 1}} \circ g} \right)^V}
\label{eq:xdef138}\end{equation}
 
This allows to define a Gaussian distribution for the state variables, we have: 
   \begin{equation}  \rho \left( g \right) = c\exp \left( { - \frac{1}{2}{x^T}Cx} \right]
\label{eq:xdef13}\end{equation}     
where:  $\int\limits_G {\rho (g)dg = 1} $ and 'g' is defined as: 
$g =\left({\bar a,A} \right)$, with: $\bar a \in {R^3}$ and $A \in SO(3) $. 
And where: $ \bar x \in \mathbb{R}^6 $ can be obtained as: $x = {\left( {\log g} \right)^V}$
                                                                                            
It can be shown, ~\cite{WangChirikjian:2006:ErrorPropagation}, that if we define the matrix of covariances:
\begin{equation}
\sum  = \left\{ {{\sigma _{ij}} = \int\limits_{{\Re ^6}} {{x_i}} {x_j}\rho \left( {g({x_1},{x_2},...,{x_6}} \right)d{x_1}d{x_2}...d{x_6}\left| {i,j = 1,2,...6} \right.} \right\}
   \label{eq:xdef14}\end{equation} 
  
we have: 
\begin{equation}
C = {\Sigma ^{ - 1}} \label{eq:xdef15}\end{equation}  and: 
\begin{equation}c = {\left( {8{\pi ^3}{{\left| {\det \sum } \right|}^{\frac{1}{2}}}} \right)^{ - 1}}\label{eq:xdef16}\end{equation}

 The Shannon (differential) entropy associated to such a distribution is given by:
\begin{equation}
        S\left( {\rho \left( g \right)} \right) = \log \left\{ {{{\left( {2\pi e} \right)}^{{\raise0.7ex\hbox{$n$} \!\mathord{\left/
 {\vphantom {n 2}}\right.\kern-\nulldelimiterspace}
\!\lower0.7ex\hbox{$2$}}}}{{\left| \sum  \right|}^{\frac{1}{2}}}} \right\} \label{eq:xdef17}\end{equation}  
Let's consider now a kinematic serial chain made of a series of rigid bodies identified by a set of frames. It can be shown that given n shifted frames, with 'tight' gaussian distributions, we have a closed form expression for the quantities in equation \eqref{eq:xdef17}:
\begin{equation}
   \sum {}  = C_{}^{ - 1} = \sum\limits_{i = 1}^{n - 1} {\left\{ {A{d_{{i^{ - 1}}}}\sum\nolimits_i {Ad_{{i^{ - 1}}}^T}  + \sum\nolimits_n {} } \right\}} 
   \label{eq:xdef18}\end{equation} 

with: 
   \begin{equation}
 A{d_i}\bar k = {\left( {g\left( {\sum\nolimits_{j = 1}^6 {{k_j}} {{\tilde E}_j}} \right){g^{ - 1}}} \right)^V}
 \label{eq:xdef19}\end{equation}
 
This allows us to compute the predictive information directly and in closed form and to derive a reinforcement learning rule maximizing it. At least in the previous hypotheses of 'tight' , 'concentrated' in technical language, gaussian distributions.
\section*{APPENDIX B. Expression of predictive information}

It is now possible to derive a reinforcement learning rule with reference to known methods and relation from information theory, see ~\cite{CoverThomas:2006:Elements,DelSole:2003:Predictable}. 
The expression of the predictive information in Markov hypotheses is given by:
\begin{equation}
I\left( {{X_{t + \tau }};{X_t}} \right) = \left\langle {\log \frac{{p\left( {{x_{t + \tau }},{x_t}} \right)}}{{p\left( {{x_{t + \tau }}} \right)p\left( {{x_t}} \right)}}} \right\rangle  = \left\langle {\log \frac{{p\left( {{x_{t + \tau }}\left| {{x_t}} \right.} \right)}}{{p\left( {{x_{t + \tau }}} \right)}}} \right\rangle 
\label{eq:xdef22a}\end{equation}

For Markov processes the 'predictive information' of the sensor is equal to the mutual information between the current and the next measured values.
Let's assume, then, that the state evolution can be modeled by assuming that the noise can be separated from the state vector, for each state vector component, such as:
\begin{equation}
{x_i} = N\left( {{\mu _i},{\sigma _i}} \right) \approx {x_i} + {\omega _i}
\label{eq:xdef23a}\end{equation}

with the noise in $ {\omega _i} $:
\begin{equation}
 {\omega _i} = N\left( {0,{\sigma _i}} \right),{x_i} = {\mu _i}
\label{eq:xdef24a}\end{equation}

We can then write, assuming a linear control function (a linear local approximation of the control system), the stochastic model: 
\begin{equation}
\dot X = AX + W
\label{eq:xdef25a}\end{equation}
          
where: $W = N\left( {0,\Sigma } \right)$
with $\Sigma$ given by previous equation (18).
If we assume the process is stationary, we have:
\begin{equation}
p\left( {{x_t}} \right) = p\left( {{x_{t + }}} \right) = N\left( {0,{\Sigma _v}} \right)
\label{eq:xdef26a}\end{equation}

with: 
\begin{equation}
 {\Sigma _v} = \int\limits_0^\infty  {{e^{As}}} \Sigma {e^{{A^T}s}}ds
\label{eq:xdef27a}\end{equation}

We can express the distribution of the 'future' sensor values in term of the 'current' ones, as we have:
\begin{equation}
{x_{t + \tau }} = {e^{A\tau }}{x_t} + {\eta _v}(t + \tau )
\label{eq:xdef27a}\end{equation}
          
with:
\begin{equation}
{\eta _v}\left( {t + \tau } \right) \approx N\left( {0,{\Sigma _v} - {e^{A\tau }}{\Sigma _v}{e^{{A^T}\tau }}} \right)
\label{eq:xdef29a}\end{equation}
It is then possible to express:
\begin{equation}
	p\left( {\left. {{x_{t + \tau }}} \right|{x_t}} \right) \approx N\left( {{e^{A\tau }}{x_t},{\Sigma _v} - {e^{A\tau }}{\Sigma _v}{e^{{A^T}\tau }}} \right)
\label{eq:xdef30a}\end{equation}

by applying the definition and exploiting the expression for multivariate normal distribution, we get:
\begin{equation}
\left( {{X_{t + \tau }};{X_t}} \right) =  - \frac{1}{2}\log \left( {\left| {1 - W{W^T}} \right|} \right)
\label{eq:xdef31a}\end{equation}

with:
\begin{equation}W = \Sigma _v^{ - {\raise0.7ex\hbox{$1$} \!\mathord{\left/
 {\vphantom {1 2}}\right.\kern-\nulldelimiterspace}
\!\lower0.7ex\hbox{$2$}}}{e^{A\tau }}\Sigma _v^{1/2}
\label{eq:xdef32a}\end{equation}
The control parameters are those for which we have:
\begin{equation}
\mathord{\buildrel{\lower3pt\hbox{$\scriptscriptstyle\frown$}} 
\over A}  = \arg \max \left( { - \frac{1}{2}\log \left( {\left| {1 - W{W^T}} \right|} \right)} \right)
\label{eq:xdef33a}\end{equation}

\section*{Acknowledgments}
The author thanks the reviewers and Prof. R. Pfeifer for their insightful and challenging remarks.


\clearpage
\clearpage


\clearpage

\clearpage
\clearpage

\clearpage

\clearpage

\clearpage

\clearpage

\clearpage

\clearpage

\clearpage

\end{document}